\documentclass[%
 reprint,
 amsmath,amssymb,aps,
pra,
floatfix,
]{revtex4-2}
\usepackage{lineno}
\usepackage{epsfig}
\usepackage{float}
\usepackage{xcolor, soul}
\sethlcolor{green}
\usepackage{graphicx}
\usepackage{dcolumn}
\usepackage{bm}
\usepackage[utf8]{inputenc}
\usepackage[normalem]{ulem}
\usepackage[T1]{fontenc}
\usepackage[english]{babel}

\usepackage{amsmath}
\usepackage{amsfonts}
\usepackage{mdframed}
\usepackage[absolute]{textpos}
\usepackage{colortbl}
\usepackage{array}

\usepackage[colorlinks]{hyperref}
\usepackage{color}

\begin{document}
\preprint{APS/123-QED}


\title{Envelope vector solitons in nonlinear flexible mechanical metamaterials}
\author{A.\,Demiquel}
\author{V.\,Achilleos}
\author{G.\,Theocharis}
\author{V.\,Tournat}

\affiliation{Laboratoire d’Acoustique de l’Université du Mans (LAUM), UMR 6613, Institut d’Acoustique - Graduate School (IA-GS), CNRS, Le Mans Université, France}

\date{\today}

\begin{abstract}

In this paper, we employ a combination of analytical and numerical techniques to investigate the dynamics of lattice envelope vector soliton solutions propagating within a one-dimensional chain of flexible mechanical metamaterial. To model the system, we formulate discrete equations that describe the longitudinal and rotational displacements of each individual rigid unit mass using a lump element approach. By applying the multiple-scales method in the context of a semi-discrete approximation, we derive an effective nonlinear Schrödinger equation that characterizes the evolution of rotational and slowly varying envelope waves from the aforementioned discrete motion equations. We thus show that this flexible mechanical metamaterial chain supports envelope vector solitons where the rotational component has the form of either a bright or a dark soliton. In addition, due to nonlinear coupling, the longitudinal displacement displays kink-like profiles thus forming the 2-components vector soliton. These findings, which include specific vector envelope solutions, enrich our knowledge on the nonlinear wave solutions supported by flexible mechanical metamaterials and open new possibilities for the control of nonlinear waves and vibrations.


\end{abstract}

\
\maketitle


\section{Introduction}


Nonlinear flexible mechanical metamaterials (FlexMMs) are an emerging class of engineered materials often consisting of highly deformable soft elements connected to stiffer ones \cite{bertoldi_flexible_2017}. They encompass a variety of designs such as origami \cite{filipov_origami_2015,miyazawa_topological_2022} and kirigami structures \cite{shyu_kirigami_2015,isobe_initial_2016}, assembled mechanical parts, 3D-printed multimaterials \cite{raney_printing_2015,sundaram_topology_2019}, and have been shown to exhibit "exotic functionalities, such as pattern and shape transformations in response to mechanical forces, or reprogrammability" \cite{bertoldi_flexible_2017}. Their capacity to undergo large local deformations, including local rotations, stems from the high elasticity contrasts together with their structure and naturally implies geometric non-linearity. As with other types of metamaterials, their linear properties depend on the geometry of the structure in addition to the constituent materials, so that both non-linear and linear mechanical behaviors can be tuned by modifying their structural or material parameters. Interestingly, in the context of wave control, harnessing the nonlinear properties of a metamaterial is particularly novel, since the majority of reported results have focused on controlling linear waves by managing dispersive effects.

Despite linear wave metamaterials constitute the vast majority of studied wave control strategies, a number of nonlinear wave effects have been studied and revealed in such flexible mechanical metamaterials \cite{deng_nonlinear_2021}, including pulse vector solitons \cite{deng_elastic_2017,deng_metamaterials_2018,deng_anomalous_2019}, rarefaction solitary waves \cite{herbold_propagation_2013,deng_propagation_2019}, transition waves and topological solitons through bistable structures for example \cite{jin_guided_2020,zareei_harnessing_2020,yasuda_transition_2020}, and more recently the manifestation of modulation instability (MI) \cite{demiquel_modulation_2023}.

However, to our knowledge, envelope solitons (bright and dark solitons) or breathers, have not been reported in FlexMM. Bright and dark solitons are solutions of the universal nonlinear Schrödinger equation (NLS), and result from the complex interplay between the dispersion and nonlinearity properties of a medium \cite{ablowitz_discrete_2004,peyrard_physics_2010}. On the one hand bright solitons \cite{silberberg_collapse_1990}, are characterized by their ability to maintain a focused intensity peak during propagation. As such, these wave objects have practical applications in optical communication systems via nonlinear optical fibers, contributing to the stability and robustness of information transmission \cite{hasegawa_transmission_1973,mollenauer_experimental_1980}. On the other hand, dark solitons, which manifest themselves as stable and localized intensity drops in a wave train, have been studied in various physical contexts, including Bose-Einstein condensates in ultracold atomic gases \cite{pitaevskii_bose-einstein_2016}, water tank experiments \cite{chabchoub_experimental_2013} and optics \cite{emplit_picosecond_1987,krokel_dark-pulse_1988,weiner_experimental_1988}.

In terms of applications, both bright and dark solitons find utility in fields such as signal processing, optical communications, and ultrafast optics. Thus, we believe that the study of bright and dark solitons in FlexMMs will be useful in controlling large amplitude vibrations. We also expect to observe nonlinear wave phenomena in FlexMMs not yet reported in mechanics or for other wave fields.
Indeed, while bright and dark solitons are mostly associated with optical systems and cold atoms, recent research has expanded their relevance to mechanical devices. For example, researchers have explored soliton-like phenomena in structures such as granular chains or phononic crystals \cite{chong_dark_2013,chong_damped-driven_2014}. These granular solitons show potential applications in shock absorption and energy transfer mechanisms \cite{nesterenko_waves_2018}. In addition, dark solitons have been studied in the context of acoustic waves with acoustic transmission lines \cite{zhang_dark_2018}, leading to the development of novel devices for sound manipulation and waveguiding \cite{deymier_nonlinear_2013}. The interdisciplinary study of solitons in mechanical systems reflects a growing interest and understanding of their universal properties. The applications of bright and dark solitons in mechanical devices continue to expand, paving the way for innovations in fields such as acoustics, wave engineering, and materials science.
\newline
\\
The main objective of this paper is to study the bright and dark soliton solutions of the NLS equation as lattice envelope vector solitons in the nonlinear FlexMM context. The paper is structured as follows. In Sec.\;\ref{sec: Lumped element} we present the nonlinear  discrete lump model which was found to be relevant for describing the dynamical equations of FlexMMs. In section\;\ref{Modulated_waves_in_FlexMM_part} we derive an effective NLS equation (eNLS) for the slowly varying envelope of waves of the
rotational degree of freedom (DOF) in the semi-discrete approximation using asymptotic expansion and multiple-scale methods, from the discrete equations of motion of the system. Finally in sections\;\ref{Bright_soliton} and \;\ref{Dark_Soliton}, the existence and dynamics of bright and dark envelope vector solitons is investigated respectively.



\section{Lumped element approach}
\label{sec: Lumped element}
\subsection{\label{sec:level2} Problem position and modeling of the structure }
\begin{figure*}[ht!]
\begin{center}
\includegraphics[width=0.9\textwidth]{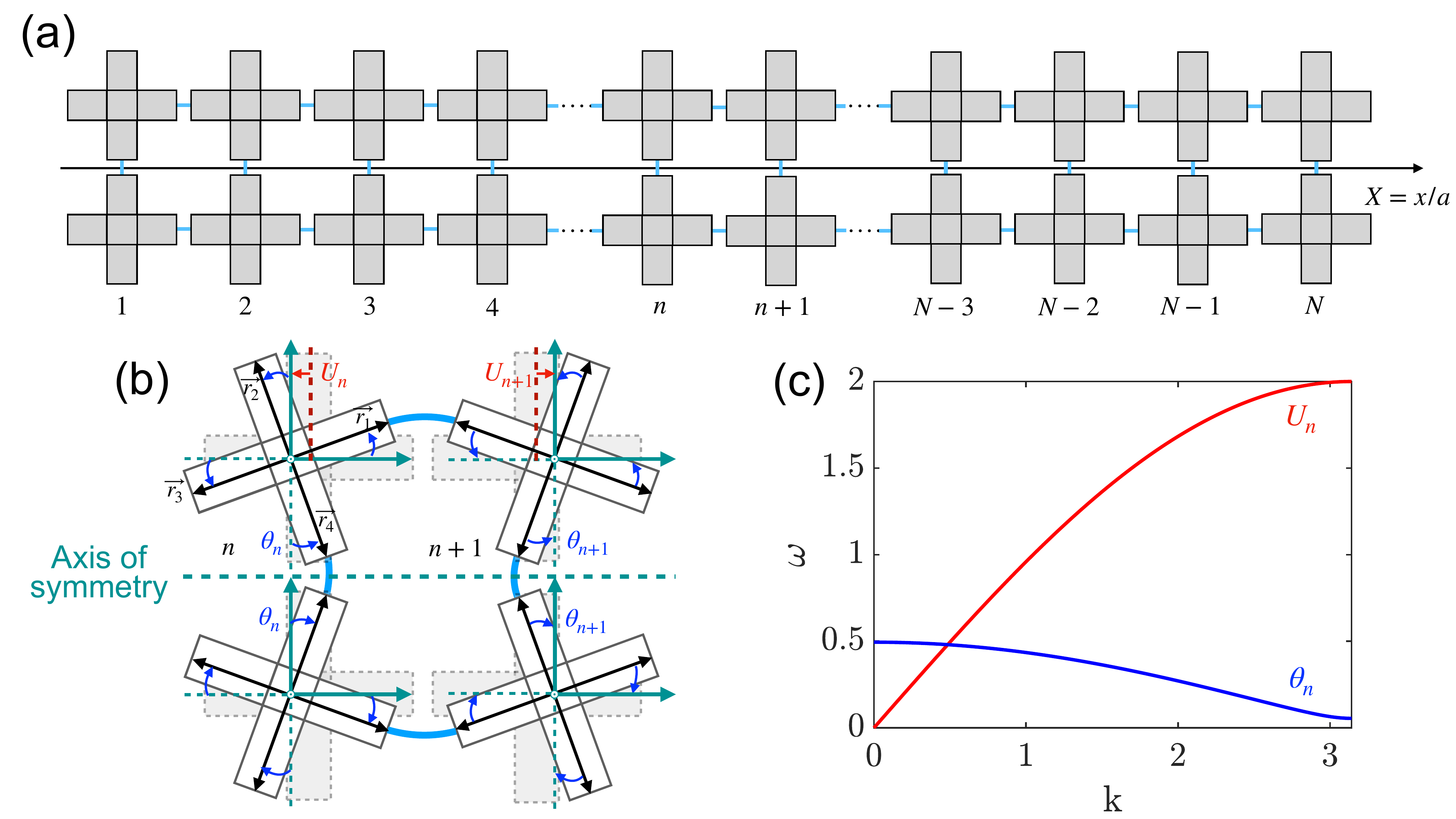}
	\caption{\label{Angle_definition} (a) Sketch of the FlexMM under consideration.
 The structure consists of two rows of rigid mass units (gray crosses) connected by elastic links (thick, blue lines) extending along the normalized X-direction ($X=x/a$) with a periodic arrangement.
The rigid units are characterized by a mass $m$ and a moment of inertia $J$ using a normalization (cf.~Eq.~(\ref{Full_discrete:sub_Theta})). The inertia of the particle can be defined by a single coefficient $\alpha$. The elastic connectors are characterized by effective stiffnesses (normalized to the longitudinal spring $k_l$): $K_s$ and $K_\theta$. 
We consider symmetric movements relative to the horizontal axis of symmetry between the two lines. (b) Displacements of the $n$ and $n+1$ particles from the equilibrium position, the mass units can rotate $\theta$ and longitudinally translate $U$. (c) Dispersion relation cf.~Eq.~(\ref{DR}) of the corresponding structure using coefficients found in the literature \cite{deng_metamaterials_2018,guo_nonlinear_2018}: $\alpha= 1.815$, $K_s = 0.01851$ and $K_{\theta} =1.534.10^{-4}$.}
	\vspace{-0.6cm}
	\label{disp_rel_example}
\end{center}
\end{figure*}

The family of FlexMM that we consider in this work consists of rigid particles (in the shape of crosses) connected to their nearest neighbors by elastic connectors and periodically arranged in a chain of two rows and $N$ columns, see Fig.\,\ref{Angle_definition}(a). 
This type of structure is inspired by the FlexMM studied experimentally in ref.~\cite{deng_metamaterials_2018}. There, the particles are constructed with Lego\textsuperscript{\textregistered} bricks and the elastic connectors are made with highly flexible plastic films.

To model this structure in the low-frequency regime, we adopt the lumped-element approach. We consider the particles as rigid, characterized by their mass $m$ and their moment of inertia $J$, 
while the elastic connectors are modeled as three massless springs; a longitudinal spring with stiffness $k_l$, a shear spring with shear stiffness $k_s$, and a bending spring with bending stiffness $k_{\theta}$. We focus on in-plane motion, so in general each particle has three DOFs, one rotation (around the z axis) and two displacements (one in the longitudinal direction along x and one in the transverse direction along y).
We also consider only symmetric motion relative to the symmetry axis, see Fig.\,\ref{Angle_definition}(b).

Similar to Ref.~\cite{deng_metamaterials_2018}, we consider two DOFs, a longitudinal displacement $u$ and a rotational motion $\theta$. This means that we constrain the particles not to move along the y-axis. Ignoring the transverse motion may also be valid even without this forced constraint. In fact, for the structure of Ref.~\cite{deng_metamaterials_2018} it was shown numerically and experimentally that during the soliton propagation \cite{deng_metamaterials_2018,deng_elastic_2017}, ignoring the transversal displacement is a reasonable assumption, since the transversal displacement amplitude was experimentally found to be an order of magnitude smaller than the longitudinal one. 
The sign of the rotation angles is considered positive in the trigonometric direction.  In previous works \cite{deng_metamaterials_2018,demiquel_modulation_2023}, the angle is defined as a positive-negative alternation from one cell to the next, which is a different point of view but does not change the physics. The choice made here makes the analysis easier, since we are studying modulated waves. 

\subsection{Motion equations of the system}
Based on the previous assumptions, we can establish the governing equations of a one-dimensional chain starting with perfect initial alignment, resulting in a static angle of zero. Considering a single element (a cross), the position of its extremities are,

\begin{equation}
\begin{split}
\mathbf{r_1} &= \begin{pmatrix}
l\cos\theta_n \\[3mm]
 l\sin\theta_n\\[3mm]
\end{pmatrix}\hspace{1cm}
\mathbf{r_2} = \begin{pmatrix}
- l\sin\theta_n \\[3mm]
l \cos\theta_n\\[3mm]
\end{pmatrix}\\
\mathbf{r_3} &= \begin{pmatrix}
-l\cos\theta_n \\[3mm]
-l\sin\theta_n\\[3mm]
\end{pmatrix}\hspace{0.7cm} 
\mathbf{r_4} = \begin{pmatrix}
l \sin\theta_n \\[3mm]
-l\cos\theta_n\\[3mm]
\end{pmatrix} 
\end{split}\,.
\end{equation} 

Thanks to the position of the vertices, the elongations of horizontally oriented springs are given by,

\begin{equation}
\begin{split}
    \mathbf{\Delta l_{n,1}}&= \mathbf{y_n}+\left[\{\mathbf{r_3}(\theta_{n+1})-\mathbf{r_3}(0)\}-\{\mathbf{r_1}(\theta_{n})-\mathbf{r_1}(0)\}\right]\\
    &=\begin{bmatrix}
u_{n+1}-u_n-l\cos\theta_{n+1}-l\cos\theta_n+2 l\\
-l\sin\theta_{n+1}-l\sin\theta_n 
\end{bmatrix}\, ,
    \end{split}
\end{equation}
while the elongation of the vertically oriented springs is neglected, due to symmetry as mentioned above,
\begin{equation}
     \mathbf{\Delta l_{n,2}} =\mathbf{0} \,.
\end{equation}

Rotational elongation can also be expressed by,
\begin{subequations}
    \begin{align}
        \Delta \theta_{n,1} &= \theta_{n+1}-\theta_{n}\, ,\\
          \Delta \theta_{n-1,1} &= \theta_{n}-\theta_{n-1}\, ,\\
        \Delta \theta_{n,2} &= 2\theta_n\, .
    \end{align}
\end{subequations}
Combining the expression of each hinge elongation, the expression for the potential energy of the system can be written,
\begin{equation}
    \mathcal{U}_{n,p}(\mathbf{\Delta l_{n,p}},\Delta \theta_{n,p})=\frac{1}{2}\lVert \mathbf{k}.\mathbf{\Delta l_{n,p}}\rVert^2+\frac{1}{2}k_\theta{\Delta \theta_{n,p}}^2 \, ,
\end{equation}
with, $ p=\{1,2\} $ and $\mathbf{k} = (\sqrt{k_l },\;\sqrt{k_s })$. \footnote{Note that this model for the elastic potential energy assumes that the elastic bonds between vertices behave physically in the following way: the bending/rotational restoring moment just depends on the relative angles between the neighboring units, the shear restoring force is proportional to the elongation of the connector projected on the axis orthogonal to the connector axis at rest (e.g. a vertical displacement difference of the vertices for a horizontal connector), and the longitudinal restoring force is proportional to the elongation of the connector projected on the axis of the connector axis at rest.  A more general model could be implemented, accounting for global rotation effects and geometrical nonlinearity associated to large rotations, but would not necessarily lead to tractable motion equations. These assumptions have been previously experimentally validated for soliton propagation in similar metamaterial chains \cite{deng_metamaterials_2018,deng_anomalous_2019,deng_nonlinear_2021}.}
\newline
\\
The Hamiltonian of the total system can then be written,
\begin{equation}
\begin{split}
     \mathcal{H}=&2\sum_{n=1}^{N}\left\{\frac{1}{2}m \dot{u_n}^2+\frac{1}{2}J\dot{\theta_n}^2\right\}+2\sum_{n=1}^{N-1}\mathcal{U}_{n,1}\left(\mathbf{\Delta l_{n,1}},\Delta \theta_{n,1}\right)\\
    &+\sum_{n=1}^{N}\mathcal{U}_{n,2}\left(\mathbf{\Delta l_{n,2}},\Delta \theta_{n,2}\right)\, ,
\end{split}
\end{equation}
from which motion equations can be derived, assuming symmetry of the motions relative to the horizontal symmetry axis of the chain, 

\begin{subequations}
    \begin{align}
    m\ddot{u}_n &= -\frac{1}{2}\frac{\partial \mathcal{H}}{\partial u_n }=-\frac{\partial \mathcal{U}_{n-1,1}}{\partial u_n}-\frac{\partial \mathcal{U}_{n,1}}{\partial u_n}\, , \\
    J\ddot{\theta}_n &= -\frac{1}{2}\frac{\partial \mathcal{H}}{\partial \theta_n} =-\frac{\partial \mathcal{U}_{n-1,1}}{\partial \theta_n}-\frac{\partial \mathcal{U}_{n,1}}{\partial \theta_n}-\frac{1}{2}\frac{\partial \mathcal{U}_{n,2}}{\partial \theta_n}\,.
    \end{align}
\end{subequations}
The corresponding normalized equations of motion for the $n$-th column are then written,

\begin{subequations}
      \begin{align}
     \frac{d^2U_n}{dT^2} &= U_{n+1}-2U_n+U_{n-1}-\frac{\cos\theta_{n+1}-\cos\theta_{n-1}}{2}\label{Full_discrete:sub_U} \,,\\
\frac{ 1}{\alpha^2}\frac{d^2\theta_n}{dT^2} &= K_\theta\left(\theta_{n-1}-4\theta_n+\theta_{n+1}\right)- K_s \cos\theta_n\left[\sin\theta_{n-1} \right.
\nonumber\\&
\left.+ 2\sin\theta_n+\sin\theta_{n+1}\right]- \sin\theta_n \left[2(U_{n+1}-U_{n-1})\right.
\nonumber\\&
\left.+4-\cos\theta_{n-1}-2\cos\theta_n-\cos\theta_{n+1}\right]\label{Full_discrete:sub_Theta}\,,
    \end{align}
    \label{Full_discrete}
\end{subequations}
where we have introduced the following normalized variables and parameters: the longitudinal displacement of unit $n$,  $U_n=u_n/a$, the normalized time $T=t\sqrt{k_l/m}$, an inertial parameter $\alpha =a\sqrt{m/(4J)}$, and stiffness parameters $K_{\theta} = 4 k_{\theta}/\left(k_l a^2\right)$ and $K_s = k_s/k_l$. Above, $m$ and $J$ are the  mass and the moment of inertia of the rigid units, while $a$ is the unit cell length (distance between the centers of the masses). If we compare this set of equations cf.~Eq.~ (\ref{Full_discrete}) in the one used in our previous work about modulation instability \cite{demiquel_modulation_2023}, the connection can be made by changing the signs of angles $\theta_{n\pm 1}$.

In the linear limit, 
the two motion (displacements and rotations) are decoupled, i.e. each DOF follows its own dynamics, independent of the other.  The corresponding dispersion relations are given by,
\begin{subequations}
\begin{align}
        \omega^{(1)} & =2\sin\left(\frac{k}{2}\right)\, ,\label{omega_1:subeq1}\\
        \omega^{(2)}&=\pm\sqrt{4\alpha^2(K_s-K_\theta)\cos^2\left(\frac{k}{2}\right)+6\alpha^2 K_\theta} \, .\label{omega_2:subeq2}
        \end{align}
        \label{DR}
\end{subequations}
Displayed in the figure \ref{Angle_definition}(c), the red branch corresponds to propagating longitudinal wave: $\omega^{(1)}$, exhibiting a typical monoatomic dispersion relation see Eq.~(\ref{omega_1:subeq1}). The second branch, described by Eq.~(\ref{omega_2:subeq2}), represents propagating rotational waves with an inverse Klein-Gordon type dispersion relation: $\omega^{(2)}$. Notably, this branch has an upper cutoff frequency at $\omega^{(2)}_c = \alpha \sqrt{4K_s+2 K_\theta}$.

\section{Modulated waves in FlexMM: effective NLS equation from semi-discrete approximation }
\label{Modulated_waves_in_FlexMM_part}
Below we focus on weakly nonlinear solutions and consequently substitute  the following expansions,
\begin{equation}
        \cos{\theta_n} = 1-\frac{\theta_n^2}{2}+\ldots \, , \hspace{0.5cm}
        \sin{\theta_n} =\theta_n-\frac{\theta_n^3}{6}+\ldots \, ,
\end{equation}
to Eq.~(\ref{Full_discrete}). By keeping terms up to cubic order we end up with the following set of equations of motion,

\begin{widetext}
\begin{subequations}
    \begin{align}
    \frac{d^{2} U_{n}}{ dT^{2}} =&\; U_{n+1}-2 U_{n}+U_{n-1}-\frac{\theta_{n-1}^2 -\theta_{n+1}^2}{4}\label{sub:Motion_equation_Order_3_NL_U}\, ,\\
        \frac{d^{2} \theta_{n}}{dT^{2}} =& -\alpha^2\left(K_s-K_{\theta}\right)\left(\theta_{n-1}+2 \theta_{n}+\theta_{n+1}\right)-6K_\theta\alpha^2\theta_n+\alpha^2(K_s-1)\theta_n^3-\alpha^2\frac{\theta_n}{2}\left(\theta_{n-1}^2+\theta_{n+1}^2\right)\nonumber\\
        &+\frac{\alpha^2 K_s}{6}\left(\theta_{n-1}^3+2\theta_n^3+\theta_{n+1}^3\right) +K_s\alpha^2 \frac{\theta_n^2}{2}\left(\theta_{n-1}+\theta_{n+1}\right)-2\alpha^2\theta_n(U_{n+1}-U_{n-1})\label{sub:Motion_equation_Order_3_NL_Theta} \,.
    \end{align}
    \label{Motion_equation_Order_3_NL}
\end{subequations}
\end{widetext}
In order to study modulated traveling waves we make use of the semi-discrete approximation~\cite{kivshar_modulational_1992,daumont_modulational_1997,remoissenet_low-amplitude_1986}, where a carrier wave, obeying the discrete dispersion relation, is modulated by a slowly varying envelope function treated in the continuum limit.
%
In particular, we look for solutions of the following form,
\begin{subequations}
\begin{align}
U_n &= \epsilon U_0 +\epsilon^2 U_2\nonumber\\
&=\epsilon G_{0,n}(T)+\epsilon^2(G_{2,n}(T)e^{2i\sigma_n}+G_{2,n}^*(T)e^{-2i\sigma_n}) \, ,\\
\theta_n &=\epsilon \theta_1\nonumber\\ 
    &= \epsilon(F_{1,n}(T)e^{i\sigma_n}+F_{1,n}^*(T)e^{-i\sigma_n}) \, ,
    \label{perturbative_expansion_theta}
\end{align}
    \label{perturbative_expansion}
\end{subequations}
with $\sigma_n =kn-\omega T$. In 
this ansatz, $F_{1,n}$ is the modulation of the plane wave $\theta_n$ with phase $\sigma_n$. Also, due to the quadratic terms $\sim \theta^2$ in Eq.~(\ref{Motion_equation_Order_3_NL}), in the ansatz for $U_n$ we include both a dc-term $G_{0,n}$ and a term $G_{2,n}$ oscillating with a phase $2\sigma_n$. 
Substituting Eq.~(\ref{perturbative_expansion}) into Eq.~(\ref{Motion_equation_Order_3_NL})(a) we arrive at the following equations collecting the dc in Eq.~(\ref{sub:exponential_dep_0sigma}) and $e^{2i\sigma n}$ terms in Eq.~(\ref{sub:exponential_dep_2sigma}) respectively,
\begin{widetext}
\begin{subequations}
\begin{align}
    &\epsilon\ddot{G}_{0,n}= \epsilon\left(G_{0,n-1}-2G_{0,n}+G_{0,n+1}\right)-\frac{\epsilon^2}{2}\left(|F_{1,n-1}|^2-|F_{1,n+1}|^2\right)\, ,\label{sub:exponential_dep_0sigma}\\
\nonumber\\
  &\epsilon^2\left( \ddot{G}_{2,n}-4i\omega\dot{G}_{2,n}-4\omega^2G_{2,n}\right) =  \epsilon^2 \left(G_{2,n-1}e^{-2ik}-2G_{2,n}+G_{2,n+1}e^{2ik}\right)-\epsilon^2\frac{F_{1,n- 1}^2e^{- 2ik}-F_{1,n+1}^2e^{ 2ik}}{4}\,.\label{sub:exponential_dep_2sigma}
  \end{align}
  \label{sub:U}
\end{subequations}
\end{widetext}
Similarly, substituting Eq.~(\ref{perturbative_expansion}) into Eq.~(\ref{Motion_equation_Order_3_NL})(b) we get the following equation collecting the $e^{i\sigma n}$ terms,
\newpage
\begin{widetext}
\begin{equation}
\begin{split}
&\epsilon\left[\ddot{F}_{1,n}-2i\omega\dot{F}_{1,n}-\omega^2F_{1,n}\right]=\epsilon\alpha^2(K_\theta-K_s)\left[F_{1,n-1}e^{-ik}+2F_{1,n}+F_{1,n+1}e^{ik}\right]-\epsilon 6\alpha^2K_\theta F_{1,n}\\
&+\epsilon^3 3\alpha^2(K_s-1)|F_{1,n}|^2F_{1,n}-\epsilon^3\frac{\alpha^2}{2}\left[{2F_{1,n}\left(|F_{1,n-1}|^2+|F_{1,n+1}|^2\right)}+{F_{1,n}^*(F_{1,n-1}^2e^{-2ik}+F_{1,n+1}^2e^{2ik})}\right]\\
&+\epsilon^3\frac{K_s \alpha^2}{6}\left[3|F_{1,n-1}|^2F_{1,n-1}e^{-ik}+6|F_{1,n}|^2F_{1,n}+3|F_{1,n +1}|^2F_{1,n +1}e^{ ik}\right]+\epsilon^3 \frac{K_s \alpha^2}{2}\left[{F_{1,n}^2\left(F_{1,n-1}^*e^{ ik}+F_{1,n+1}^*e^{- ik}\right)}\right.\\
&\left.+{2|F_{1,n}|^2\left(F_{1,n-1}e^{- ik}+ F_{1,n+1}e^{ ik}\right)}\right]-\epsilon^3 2\alpha^2 F_{1,n}^*\left({G_{2,n+1}e^{2ik}-G_{2,n-1}e^{-2ik}}\right)-\epsilon^2 2\alpha^2 F_{1,n}\left({G_{0,n+1}-G_{0,n-1}}\right)\,.
  \end{split}
  \label{sub:exponential_dep_1sigma}
\end{equation}
\end{widetext}
We now proceed considering that the discrete functions $W_n(T)=\{F_{1,n}(T),G_{0,n}(T),G_{2,n}(T)\}$ are varying slowly in space and time. Therefore the continuum limit approximation is applied and the above discrete functions $W_n(T)$ are replaced by  $W(X_1,X_2,\ldots,T_1,T_2,\ldots)$, where  $X_i = \epsilon ^i X$ and $T_i = \epsilon ^i T$ are slow variables with $i=1,2,\ldots$ . Note that under this approximation the slowly varying functions are independent of the fast variables $n$ and $T$.
In addition $W_{n\pm1}$, 
is computed up to order $\epsilon^2$ using Taylor expansion,
\begin{equation}
  W_{n\pm1}  =  W\pm \epsilon \frac{\partial W}{\partial X_1 }\pm \epsilon^2 \frac{\partial W}{\partial X_2}+\frac{\epsilon^2}{2}\frac{\partial^2 W}{\partial X_1^2}+\mathcal{O}(\epsilon^3) \, ,
  \label{Taylor_expansion}
\end{equation}
and the time derivation as,
\begin{equation}
   \dot{W_n}=\frac{\partial W}{\partial T} = \epsilon \frac{\partial W}{\partial T_1}+  \epsilon^2 \frac{\partial W}{\partial T_2}+\mathcal{O}(\epsilon^3) \,.
   \label{Temporal}
\end{equation}
Substituting Eqs.(\ref{Taylor_expansion}-\ref{Temporal}) into the set of Eqs.\,(\ref{sub:U}-\ref{sub:exponential_dep_1sigma}) we arrive at a system of equations at successive orders in $\epsilon$.
The lowest order in Eq.~(\ref{sub:exponential_dep_0sigma}) (analogous to $\epsilon^3$) gives us a relation between the dc-term $G_0$ and the envelope of the modulated plane wave $F_1$,
\begin{equation}
    \left(\frac{\partial^2}{\partial T_1^2} -\frac{\partial^2}{\partial X_1^2}\right)G_0 = \frac{\partial |F_1|^2}{\partial X_1}.
    \label{G0}
\end{equation}
In Eq.~(\ref{sub:exponential_dep_2sigma}) the lowest order is analogous to $\epsilon^2$ and relates $G_2$ to $F_1$ as follows,
\begin{equation}
  G_2 = \frac{i\sin(2k)}{8\left(\sin^2(k)-\omega^2\right)}F_1^2.
\end{equation}

\begin{figure}[!ht]
    \centering
\includegraphics[width=0.49\textwidth]{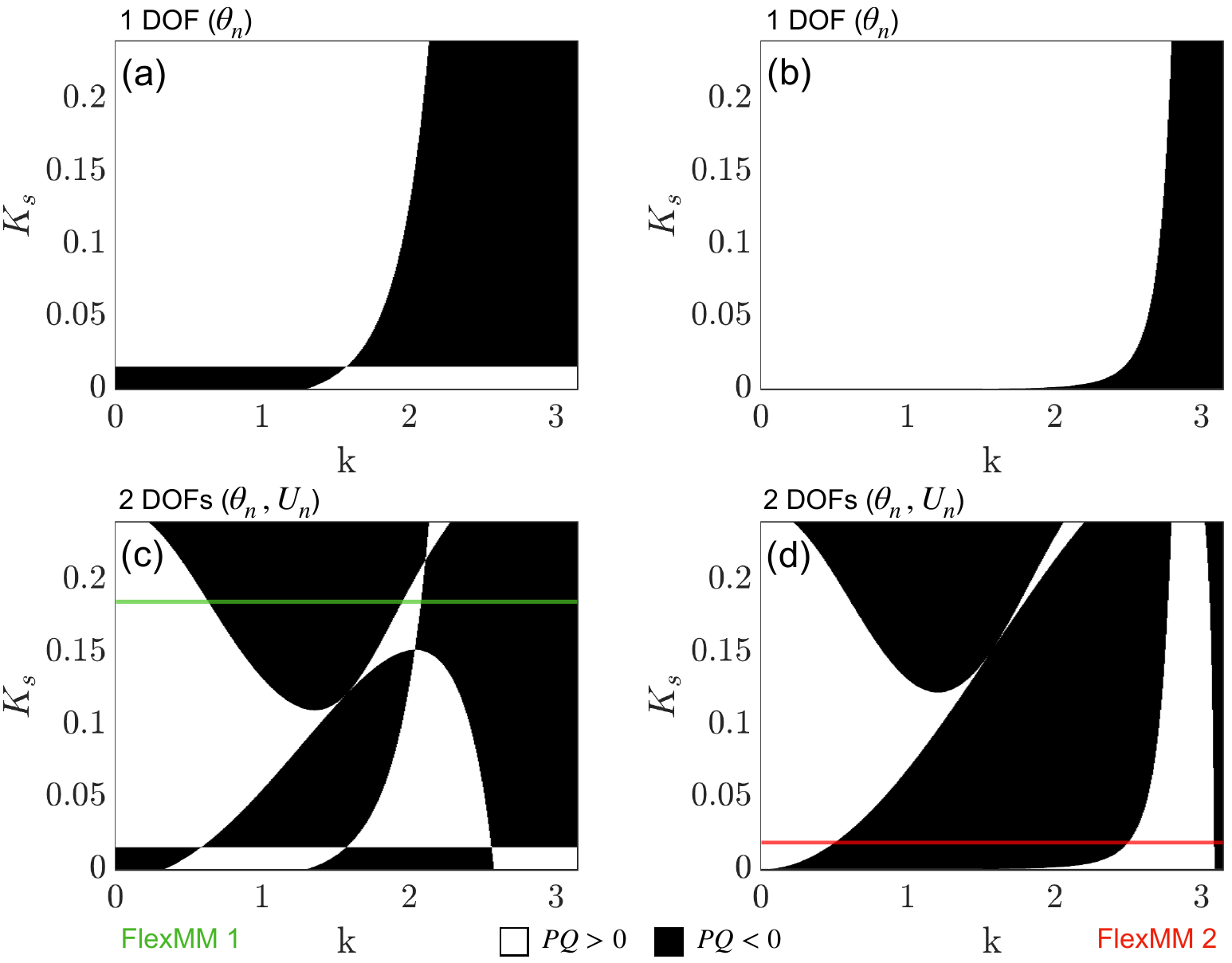}
    \caption{Sign of $PQ$ as a function of $k$ and $K_s$ for two different  $K_\theta=[1.534e-2,1.534e-4]$ respectively used in panels (a-c) and (b-d), with $\alpha=1.815$ fixed.  Panels (a-b) correspond to a configuration where particles can only rotate, while in panels (c-d) the particles can rotate and translate. The horizontal colored lines represent the parameters chosen in Secs.\,\ref{Bright_soliton}-\ref{Dark_Soliton} to study bright and dark solitons propagation along FlexMMs.}
    \label{PQ_study_3D}
\end{figure}

We now move to  Eq.~(\ref{sub:exponential_dep_1sigma}) where at order $\epsilon^1$, we recover the 
dispersion relation,
\begin{equation}
    \omega^2 = 4\alpha^2(K_s-K_\theta)\cos^2\left(\frac{k}{2}\right)+6\alpha^2 K_\theta \, ,
    \label{order_1}
\end{equation}
which corresponds to the branch of the rotational waves of the dicrete model cf.~Eq.~(\ref{omega_2:subeq2}).
At order $\epsilon^2$ we obtain the solvability condition,
\begin{equation}
      \frac{\partial F_1}{\partial T_1}+v_g\frac{\partial F_1}{\partial X_1}=0\, ,
\end{equation}
where
\begin{equation}
    v_g =-\frac{\alpha^2(K_s-K_\theta)\sin(k)}{\omega}\, ,
\end{equation}
is the group velocity corresponding to Eq.~(\ref{order_1}). Up to this order $F_1$ is linear and not coupled to $G_{0}$, $G_2$.
At the order $\epsilon^3$ we have the contribution from all fields and nonlinearity, leading to 
the following nonlinear 
 Schrödinger (NLS) equation, 
%
\begin{equation}
    i\frac{\partial F_1}{\partial \tau_2}+P\frac{\partial^2 F_1}{\partial \xi_1^2}+Q|F_1|^2F_1 =0 \, ,
    \label{NLS}
\end{equation}
in terms of the slow variables $\xi_1 = \epsilon(X - v_g T )$ and  $\tau_2= \epsilon^2 T$. $P$ and $Q$, are the dispersion and nonlinear coefficients respectively given by the following expressions,
\begin{subequations}
    \begin{align}
        P =& \frac{\alpha^2(K_\theta-K_s)\cos(k)-v_g^2}{2\omega}\label{sub:P}\, ,\\
        Q =& \frac{1}{2\omega}\left[8K_s\alpha^2\cos^2\left(\frac{k}{2}\right)-\alpha^2(5+\cos(2k))\right.\nonumber\\
        &\left.+\frac{\alpha^2\sin^2(2k)}{2\left(\sin^2(k)-\omega^2\right)}-\frac{4\alpha^2}{v_g^2-1}\right]\label{sub:Q} \, .
        \end{align}
        \label{PQ}
\end{subequations}
 We note that the last two terms in Eq.~(\ref{sub:Q}) arise due to the presence of $G_0$ and $G_2$ at order $\epsilon^3$ (in the case that we consider only rotation DOF these terms are absent), and have an important effect on the resulting eNLS properties.
\newline 
\\
 The NLS equation exhibits two distinct behaviors depending on the sign of the product $PQ$.
When $PQ>0$, it is known as focusing featuring modulational instability and bright  soliton solutions among others, while for $PQ<0$, it is referred to as defocusing with stable plane waves and dark solitons \cite{ablowitz_discrete_2004}. For our system the sign of $PQ$ is determined by the choice of the carrier wavenumber $k$ and the design characteristics of the flexible metamaterial (FlexMM) through the parameters $\alpha$,  $K_s, K_\theta$ \cite{demiquel_modulation_2023}. 
\newline
\\
In figure \ref{PQ_study_3D}, we show the sign of $PQ$ as a function of the wavenumber $k$ for the fixed inertial parameter $\alpha = 1.815$ which corresponds to the experimental setup ref.\cite{deng_nonlinear_2021}. In addition, we choose two different cases of bending stiffness, with very small values $K_\theta = 1.534e^{-2}$ for panels (a-c) and $K_\theta = 1.534e^{-4}$, typically found in flexible elastic metamaterials. To highlight the effect of coupling between the 2DOFs (rotation and longitudinal displacement) on the nature of NLS, in panels (a-b) we plot the sign of the $PQ$ product when only rotational DOF are considered, while in panels (c-d) when both DOFs are considered. Interestingly, by comparing  panels (a-b) to (c-d), we observe that the coupling between the two DOFs dramatically changes the nature of the NLS. We also observe that the variation of $K_s$ stiffness has a stronger impact in the case of 2DOFs, panels (c-d). In conclusion the nature of the eNLS crucially depends on the design characteristics of the FlexMM and the presence of 2DOFs ($U_n$ and $\theta_n$).


\begin{figure}[!ht]
    \centering   
\includegraphics[width=0.4\textwidth]{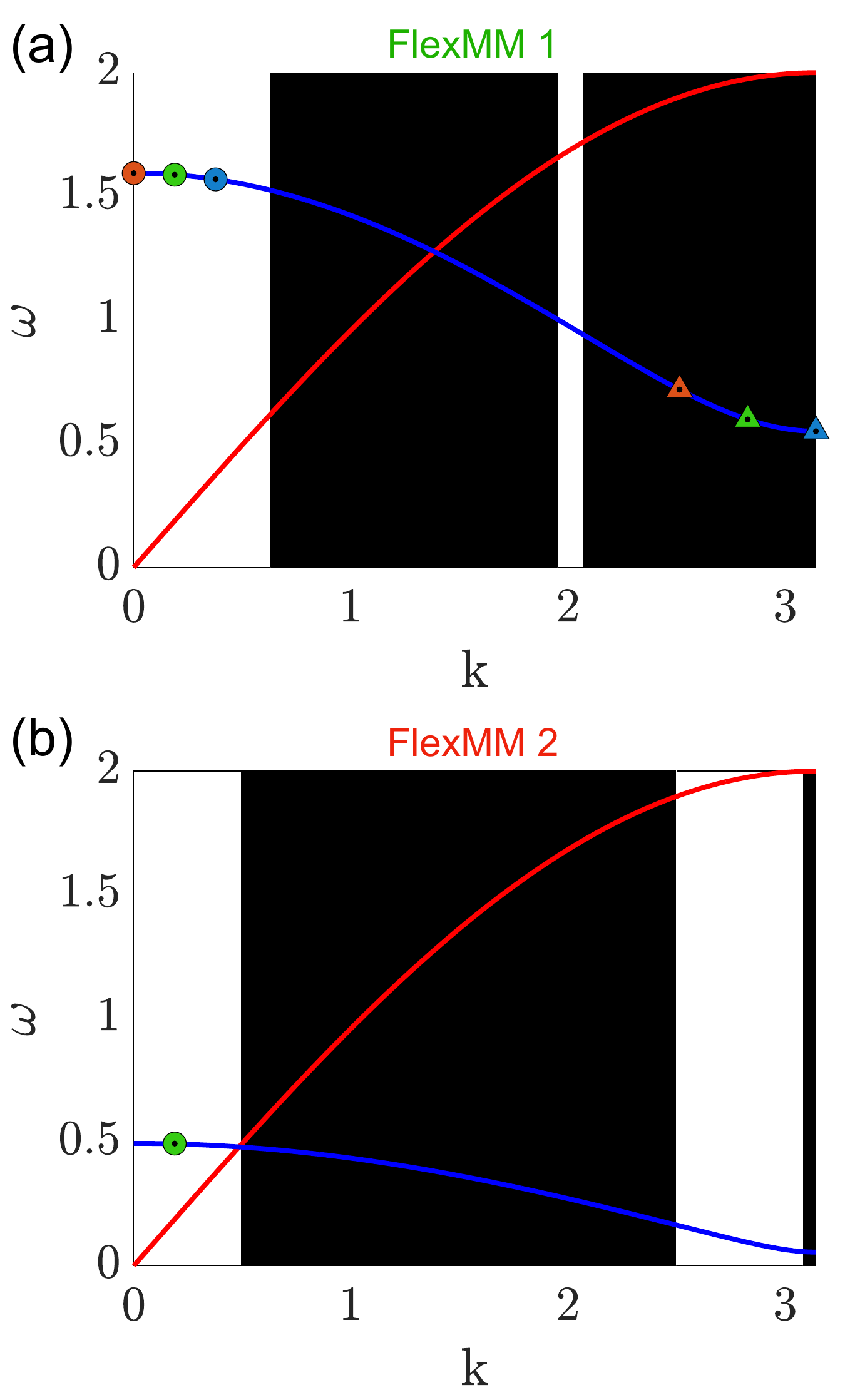}
\caption{\label{Dispersion_relation with region stable/unstable}Dispersion relations of Eq.~(\ref{Full_discrete}), derived in Eq.~(\ref{DR}). The effective NLS focusing and defocusing regions are represented by white and black areas, respectively. The colored dots correspond to the pair of $\Omega$ and $k$ used in Sec.\,\ref{Bright_soliton} to generate lattice envelope solitons. On panel (a), the dispersion relation corresponds to FlexMM 1 defined by the following set of parameters: $\alpha =1.815$, $K_s =0.1851$, $K_{\theta} =1.534\text{e}^{-2}$. For panel (b), the FlexMM 2 parameters are: $\alpha =1.815$, $K_s=0.01851$, $K_{\theta}=1.534\text{e}^{-4}$.}
    \end{figure}

For the rest of the paper we will focus on two particular designs of FlexMM corresponding to the green and red lines cf.~Fig.\,\ref{PQ_study_3D}, respectly called FlexMM 1 and FlexMM 2. 
In Fig.\,\ref{Dispersion_relation with region stable/unstable}, we show the sign of $PQ$ (as given by Eq. (\ref{PQ})) together with the  dispersion relation Eq.~(\ref{DR}) corresponding to the two FlexMM configurations mentioned above.

\section{Bright envelope vector solitons} 
\label{Bright_soliton}
\subsection{Theoretical prediction}
The focusing nonlinear Schrödinger equation, Eq.~(\ref{NLS}) with $PQ>0$, admits the following bright soliton solution \cite{kivshar_optical_2003} ,
\begin{equation}
    F_1(\xi_1,\tau_2) = A_0 \text{sech} \left[\frac{1}{L_e}(\xi_1-c\tau_2)\right]e^{ic\xi_1+iQ\left(\frac{A_0^2-c^2}{2}\right)\tau_2} \,,
     \label{F1}
\end{equation}
where
\begin{equation}
    Le = \frac{1}{A_0}\sqrt{\frac{2P}{Q}} \,,
    \label{Le}
\end{equation}
is the width, $A_0$ the amplitude and $c$ the velocity of the soliton at the co-moving frame coordinate system $\xi=X-v_gT$.



%

Using Eq.~(\ref{perturbative_expansion}), the rotation $\theta_1$ for $c=0$ is found to be
\begin{equation}
              \theta_1(X,T)=  2  A_0 \text{sech} \left[\frac{\epsilon}{Le} (X-X_0- v_g  T)\right]\cos[kx-\Omega T] \, .
          \label{Analy_theta}
\end{equation}
The angular frequency of  the carrier wave, 
   \begin{equation}
        \Omega= \omega^{(2)}-\epsilon^2\frac{Q A_0^2}{2},
        \label{capital_omega}
   \end{equation}
has been shifted at order $\epsilon^2$, in comparison to the linear dispersion relation $\omega^{(2)}$ due to nonlinearity. 
Depending on the sign of $Q$, the shift can occur above or below the linear branch. 

The combination of Eqs.~(\ref{F1}) and (\ref{G0}) gives the following expression for the dc-term,
\begin{equation}
        U_0(X,T) =\frac{A_0^2 L_e}{v_g^2-1} \tanh\left[\frac{\epsilon (X-X_0)-\epsilon v_g T}{L_e}\right]\, .
        \label{Analy_U}
\end{equation}
\\

Equations (\ref{Analy_theta}) and (\ref{Analy_U}) constitute a polarized envelope nonlinear wave solution of 
Eq.~(\ref{Full_discrete}) which is propagating with a common velocity determined by the spatial frequency of the carrier wave, defined as $v_g = \frac{d\omega^{(2)}}{dk}=-\frac{\alpha^2(K_s-K_{\theta})\sin{k}}{\omega^{(2)}}$.
From now on, we refer to is as bright envelope vector soliton (BEVS).

Note that physically this shape for the $U$ field corresponds to a longitudinal contraction of the chain around the maximum of rotating sites during the propagation of the BEVS.



\subsection{Bright Envelope vector soliton propagation in FlexMM}
\label{part:BEVS}

Direct numerical simulations of the discrete set of equations (\ref{Full_discrete}) are employed to validate our analytical predictions.
The system (\ref{Full_discrete}) is solved using a fourth-order Runge-Kutta iterative integration scheme for a total of $N=1000$ sites, with free boundary conditions at both ends.
The results presented in section \ref{Bright_soliton} were obtained by performing the integration for a duration of eight nonlinear times : $t_f = 8 T_{NL} $\cite{remoissenet_waves_1999}. $T_{NL}$ is based on the initial condition (IC) amplitude $A_0$, the system nonlinearity $Q$ 
and the carrier wave number $k$ cf.~ Eq. (\ref{PQ}). The relationship between $T_{NL}$, $A_0$ and $Q$ is given by,
\begin{equation}
        T_{NL}= \frac{1}{\epsilon^2 |Q|A_0^2} \, .
        \label{NL_length}
   \end{equation}
The initial conditions are taken to be,
    \begin{equation}
\begin{split} &\theta(X,0)= \epsilon \theta_1(X,0)\, ,\hspace{0.5cm}
\dot \theta(X,0) = \epsilon\dot \theta_1(X,0)\, , 
\end{split}
     \label{IC_T}
      \end{equation}
  \begin{equation}
\begin{split} &U(X,0) = \epsilon U_0(X,0) \, ,\hspace{0.5cm}
\dot U(X,0) = \epsilon \dot U_0(X,0)\, ,
     \end{split} 
     \label{IC_U}
       \end{equation}
using Eqs.~(\ref{Analy_theta}-\ref{Analy_U}) with $X_0 = N/2= 500$. The initial amplitude $A_0$ chosen for the next simulations is defined as $A_0=A\sqrt{\frac{2P}{Q}}$ with $A=15$. When $P =1/2$ and $Q = 1$, $A_0 = A$ is the amplitude of the bright soliton of the normalized NLS equation.
Defining the amplitude $A_0$ as a function of $P$ and $Q$ implies, from the perspective of the NLS model, that the bright soliton initial conditions for FlexMM1 and FlexMM2, represented in Figs.(\ref{fig:Bright_3D}-\ref{Bright_lines}-\ref{fig:Bright_3D_Lego}), are the same.

%

\subsubsection{Nonlinear dynamics of FlexMM 1}
\begin{figure}[ht!]
\includegraphics[width=0.48\textwidth]{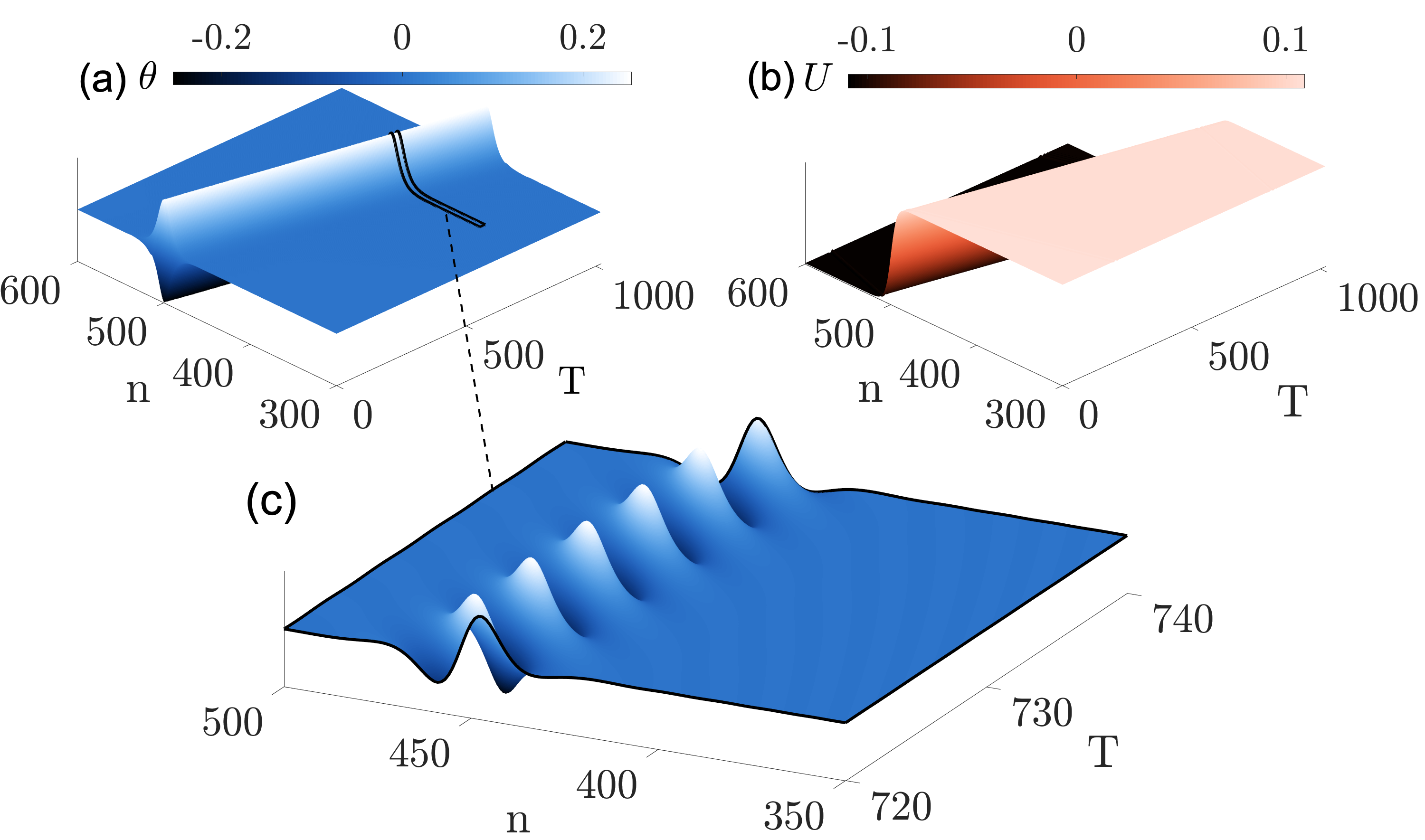}
\caption{\label{fig:Bright_3D} Evolution in time ($T$) of the amplitudes of the rotational (a-c) and longitudinal (b) displacements along the chain ($n$). The results correspond to a FlexMM (FlexMM 1) defined by the following set of parameters: $\alpha =1.815$, $K_s =0.1851$, $K_{\theta} =1.534\text{e}^{-2}$. The initial condition corresponds to a BEVS with $k=0.1885$ and $A = 15$, and a perturbation of $\epsilon = 0.01$.
}
\end{figure}

We start by studying the first FlexMM structure characterized by the dispersion relation shown in Fig. \ref{Dispersion_relation with region stable/unstable}(a), 
where as one can see, the upper cut-off frequencies of the two branches are close $\omega^{(2)}_c \approx \omega^{(1)}_c=2 $. We will focus on the small $k$ region where the effective NLS is focusing and thus BEVS are predicted. 
In Fig. \ref{fig:Bright_3D}, we show the nonlinear dynamics of an initial condition with a BEVS with $k=0.1885$, corresponding to the green circle point of Fig. \ref{Dispersion_relation with region stable/unstable}(a).
The dynamics confirm that indeed the IC evolves as a BEVS and propagates with a constant velocity keeping its shape undistorted in the form of an envelope for rotations $\theta$ (Fig. \ref{fig:Bright_3D}(a),(c)) and a kink for displacement $U$  (Fig. \ref{fig:Bright_3D}(b)).

\begin{figure}
\includegraphics[width=0.48\textwidth]{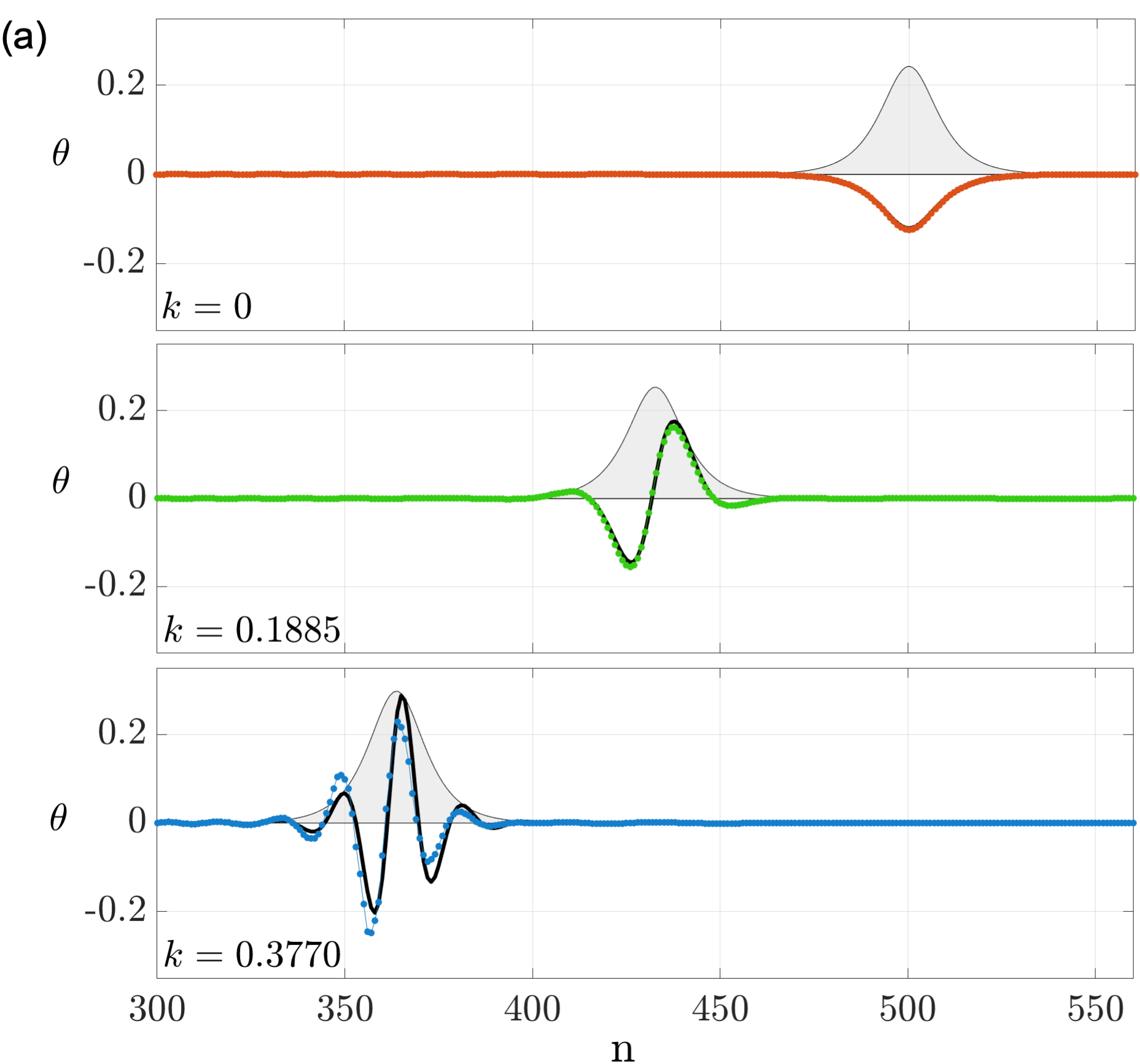}
\includegraphics[width=0.475\textwidth]{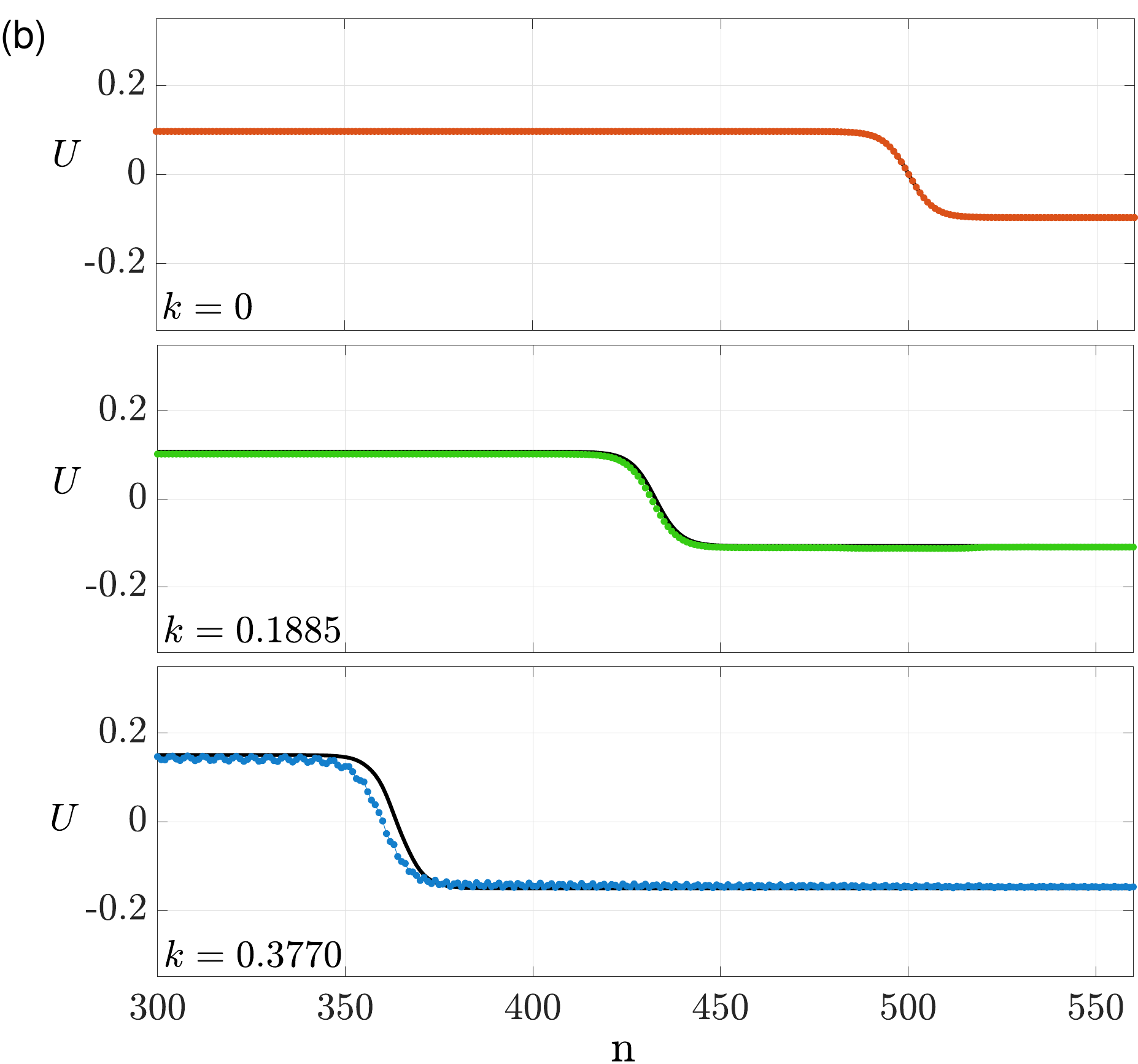}
\caption{\label{Bright_lines}Rotational and longitudinal displacement amplitudes along the chain ($n$) at final time $t_f =8 T_{NL}$.
The initial conditions correspond to BEVS with  $k=0$ in orange , $k=0.1885$ in green and $k=0.3770$ in blue, and an amplitude of
$A = 15$, and a perturbation of $\epsilon = 0.01$.
}
\end{figure}

For a more systematic study, we have performed numerical simulations of BEVS with different wavenumbers, within the focusing NLS region.
In figure \ref{Bright_lines} (a-b), we show the solution profile at 
the final time $t_f=8 T_{NL}$ for three cases corresponding to the three circles (orange, green and blue) in Fig.\,\ref{Dispersion_relation with region stable/unstable}(a). Superimposed are the theoretical solutions (black lines) given by Eqs.(\ref{Analy_theta}-\ref{Analy_U}).

As expected, the orange one ($k=0$) remains centered at $X_0$ due to $v_g=0$, see dispersion relation curves Fig.\,\ref{Dispersion_relation with region stable/unstable}(a). The other two ($k=\{0.1885, 0.3770\}$)
move in the left direction due to negative group velocities, with different velocities.
Overall, analytical predictions and simulation results are in good agreement. The BEVS predicted by the NLS bright soliton 
show a robust behavior after eight nonlinear times, confirming the validity of the effective NLS. 
This is a first signature that BEVs exists and propagate through the lattice. 
For the three cases final times of integration are close, $t_f(k = 0) \approx 1013$, $t_f (k = 0.1885)\approx 1020$, $t_f(k=0.3770)\approx 1031$.

Another signature of the BEVS propagating through the lattice can be extracted from the nonlinear dispersion relation (NDR).
As it was shown in \cite{tikan_nonlinear_2022,leisman_effective_2019}, bright solitons correspond to straight lines in the NDR.
To see this, let us use the space-time double Fourier transformation,
\begin{equation} 
     \tilde {\theta}_1(\omega,k) = \int_{-\infty}^{+\infty}\int_{-\infty}^{+\infty} \theta_1(X,T)e^{-ikX}e^{+i\omega T}\text{d}X\text{d}T,
\end{equation}
using as $\theta_1(X,T)$ the BEVS solution, namely Eq.~(\ref{Analy_theta}), at a chosen $k_s$ and $\omega^{(2)}(k_s)$. We obtain that,
\begin{equation}
\begin{split}
        \epsilon \tilde\theta_1(\omega,k)= & 2\pi^2A_0L_e\text{sech}\left(\frac{\pi L_e(k-k_s)}{2\epsilon}\right)\text{e}^{-iX_0(k-k_s)}\\
    &\times \delta(\omega_s-\frac{Q\epsilon^2A_0^2}{2}+v_g(k-k_s)-\omega)\\
    & + 2\pi^2A_0L_e\text{sech}\left(\frac{\pi L_e(k+k_s)}{2\epsilon}\right)\text{e}^{-iX_0(k+k_s)}\\
    &\times \delta(-\omega_s+\frac{Q\epsilon^2A_0^2}{2}+v_g(k+ k_s)-\omega).\\
    \end{split}
    \label{dirac}
\end{equation}
%
Using the Dirac function property, the value of $\delta(\omega_s-\frac{Q\epsilon^2A_0^2}{2}+v_g(k-k_s)-\omega)$ and therefore $\tilde \theta_1$ is 0 except for points on the line,
\begin{equation}
\omega = v_g k+\left(\omega_s -v_g k_s-\epsilon^2\frac{Q A_0^2}{2}\right)\,.
\label{NDR_soliton}
\end{equation}
Eq.~(\ref{NDR_soliton}) is the NDR of the BEVS which indeed is a straight line in the $\omega-k$ diagram.

\begin{figure}[ht!]
    \centering   
\includegraphics[width=0.48\textwidth]{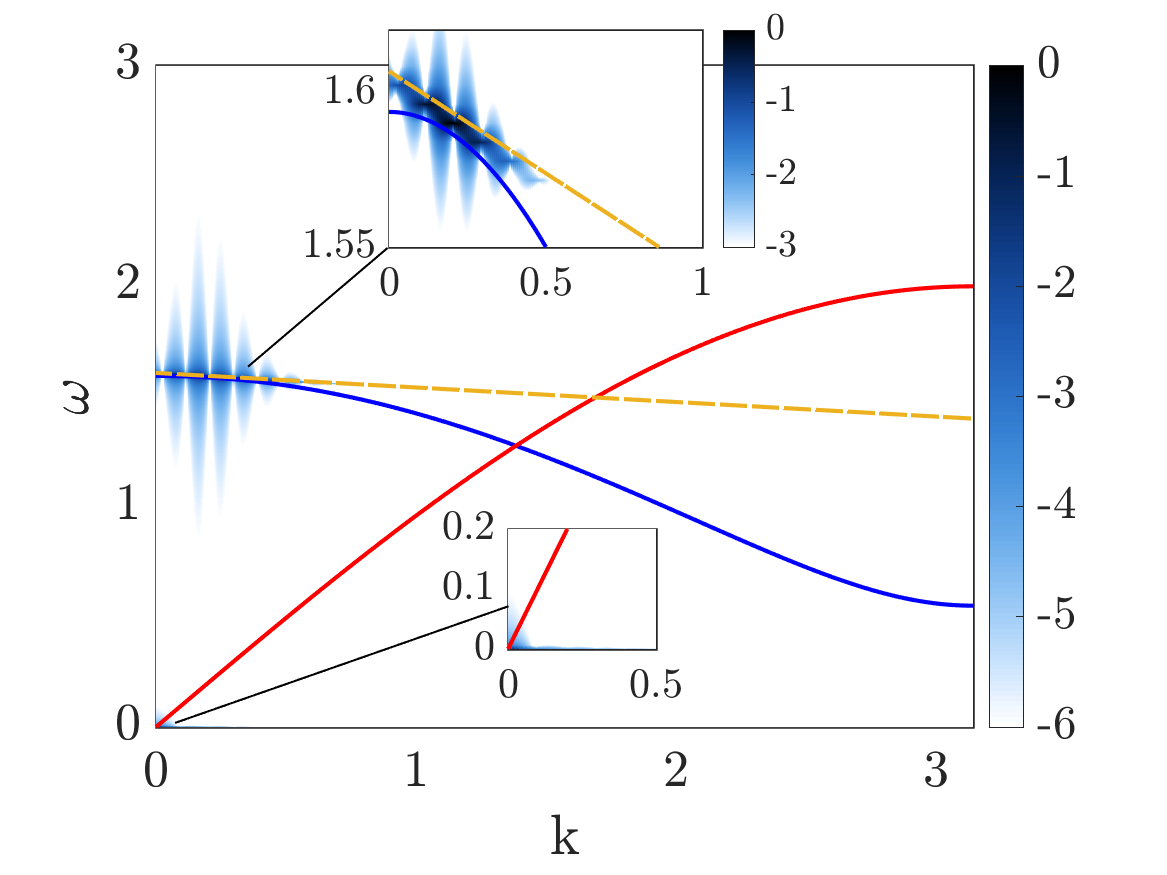}
\caption{\label{DR_white_and_zoom} Numerical representation of the nonlinear dispersion relation of FlexMM 1 from its dynamics, represented in Fig.\, \ref{fig:Bright_3D} using a normalized sum of the 2D-FFTs of the $\theta$ and $U$ components. The red and blue curves denote the linear dispersion relation (see Fig.\,\ref{Dispersion_relation with region stable/unstable}(a)), while the yellow line denotes NDR of the soliton described in Eq.~(\ref{NDR_soliton}). The color bar represents the $\tilde\psi(\omega,k)$, in $log$ scale.}
\end{figure}

Now, using the spatio-temporal dynamics, Fig.\,\ref{fig:Bright_3D}, we calculate the 
double FFT (in space and time). Since the lattice has two fields, $\theta_n(T)$
and $U_n(T)$, we apply the space-time double Fourier transform in both fields to obtain   $\tilde{\theta}(\omega,k)$ and $\tilde{U}(\omega,k)$.
In Fig.\,\ref{DR_white_and_zoom}, we represented the normalized sum of the double FFT in space and time, in $log$ scale,
\begin{equation}
\tilde\psi(\omega,k)= \left|\frac{\tilde\theta(\omega,k)}{\tilde\theta_{max}}\right|+\left|\frac{\tilde U(\omega,k)}{\tilde U_{max}}\right|\,.
\end{equation}
As it is seen, a large amount of the 2D FFT (blue color gradient) closely matches the estimate line provided by Eq.~\ref{NDR_soliton}, that is tangent to $\omega^{(2)}$ (yellow line) at point $(\Omega_s, k_s )$. This corresponds to the NDR of a bright soliton. An upper shift, of the order $\epsilon^2$ cf.~Eq.~(\ref{capital_omega}), compared  to the linear dispersion relation $\omega^{(2)}$ is visible. 
In addition, low frequency components around $k=0$ are observable, on the $\omega^{(1)} $curve . This corresponds to the dc-component of $U$ of the BEVS solution.  



\subsubsection{Nonlinear dynamics of FlexMM 2}
We now consider the FlexMM 2 configuration. Namely, the one that follows the dispersion relation shown in Fig.\,\ref{Dispersion_relation with region stable/unstable} (b) and which corresponds to the set of parameters used in experiments of \cite{deng_metamaterials_2018,guo_nonlinear_2018}.
As one can see in Fig.\,\ref{Dispersion_relation with region stable/unstable} (b), this FlexMM 
supports rotational modes (blue curve) with frequencies that are much lower than those of the translational modes  
(red curve). In particular, 
the upper cutoff frequencies of the two branches are such that $\omega^{(2)}_c \ll \omega^{(1)}_c=2 $.

\begin{figure}[ht!]
\includegraphics[width=0.4\textwidth]{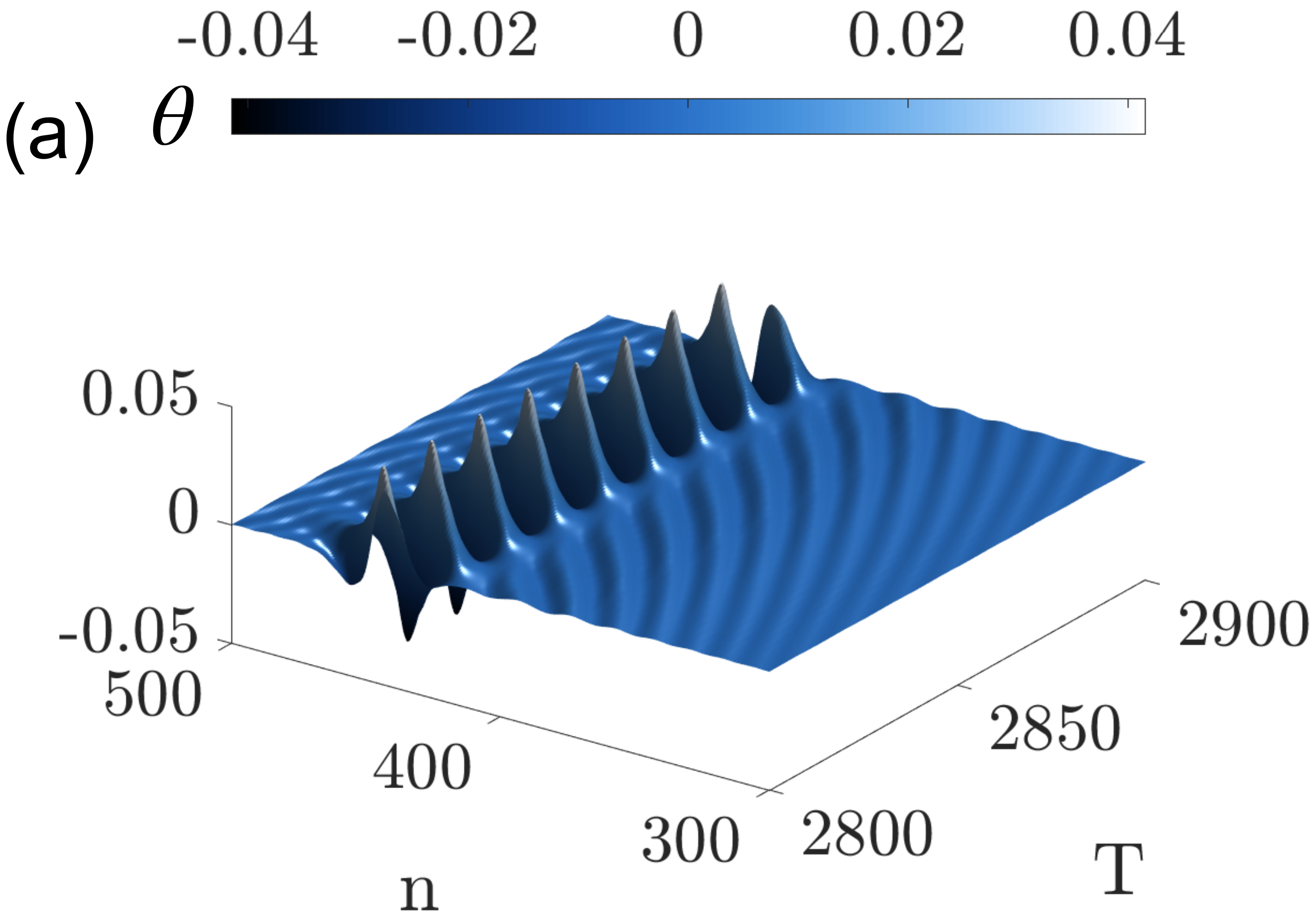}
\includegraphics[width=0.4\textwidth]{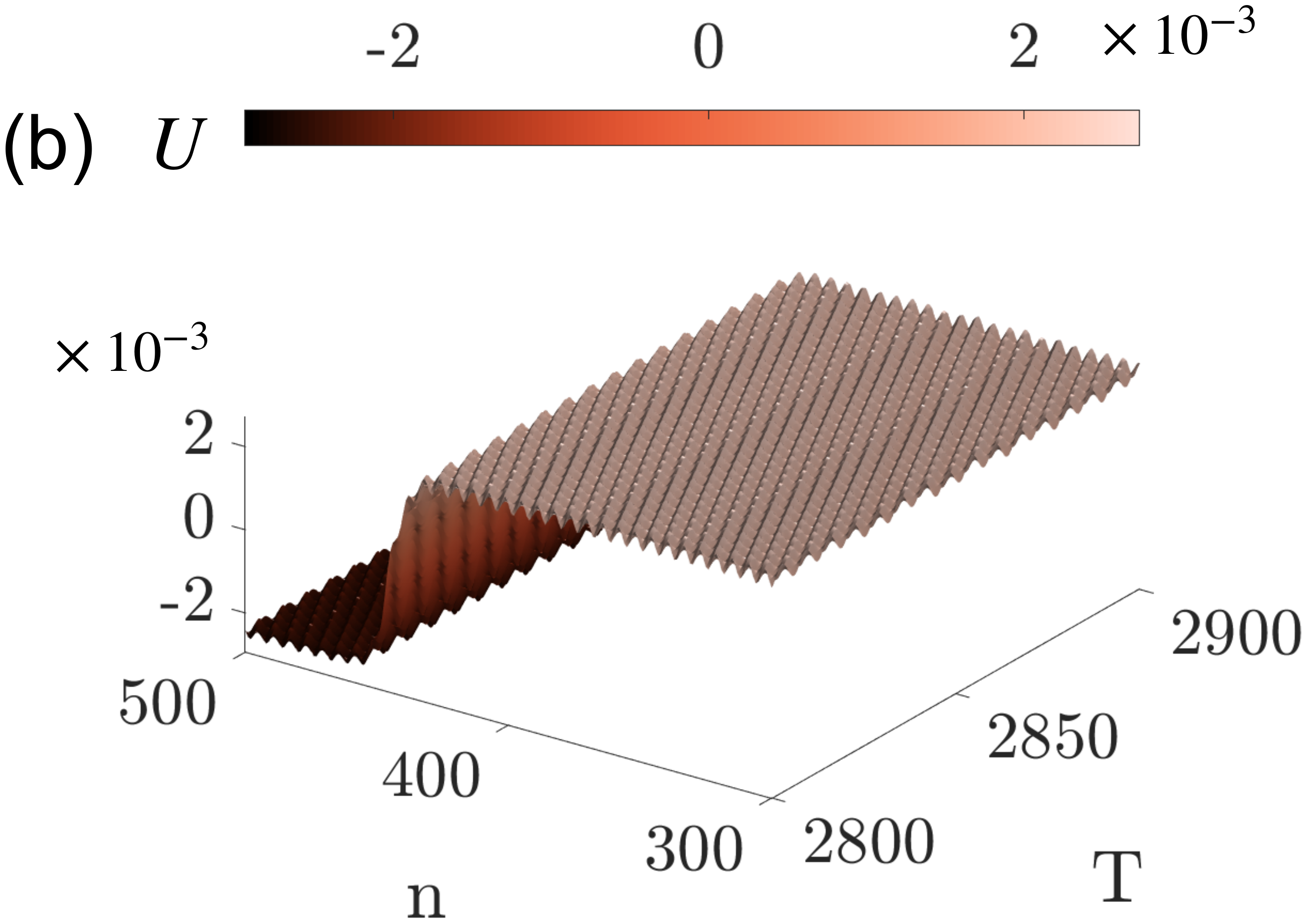}
\caption{\label{fig:Bright_3D_Lego}  Evolution in time ($T$), for a duration between $T$=$[2800 ;2900]$, of the amplitudes of rotational (a) and longitudinal (b) displacements along the chain ($n$), zoomed  between $n=[300 ;500]$. The results correspond to a FlexMM (FlexMM 2) defined by the following set of parameters: $\alpha =1.815$, $K_s =0.01851$, $K_{\theta} =1.534\text{e}^{-4}$. The bright soliton is generated by the initial conditions expressed in Eqs.\,(\ref{IC_T}-\ref{IC_U}) for a spatial frequency of $k=0.1885$, an amplitude of $A=15$, and a perturbation of $\epsilon = 0.01$. }
\end{figure}

\begin{figure}[ht!]
    \centering   
\includegraphics[width=0.48\textwidth]{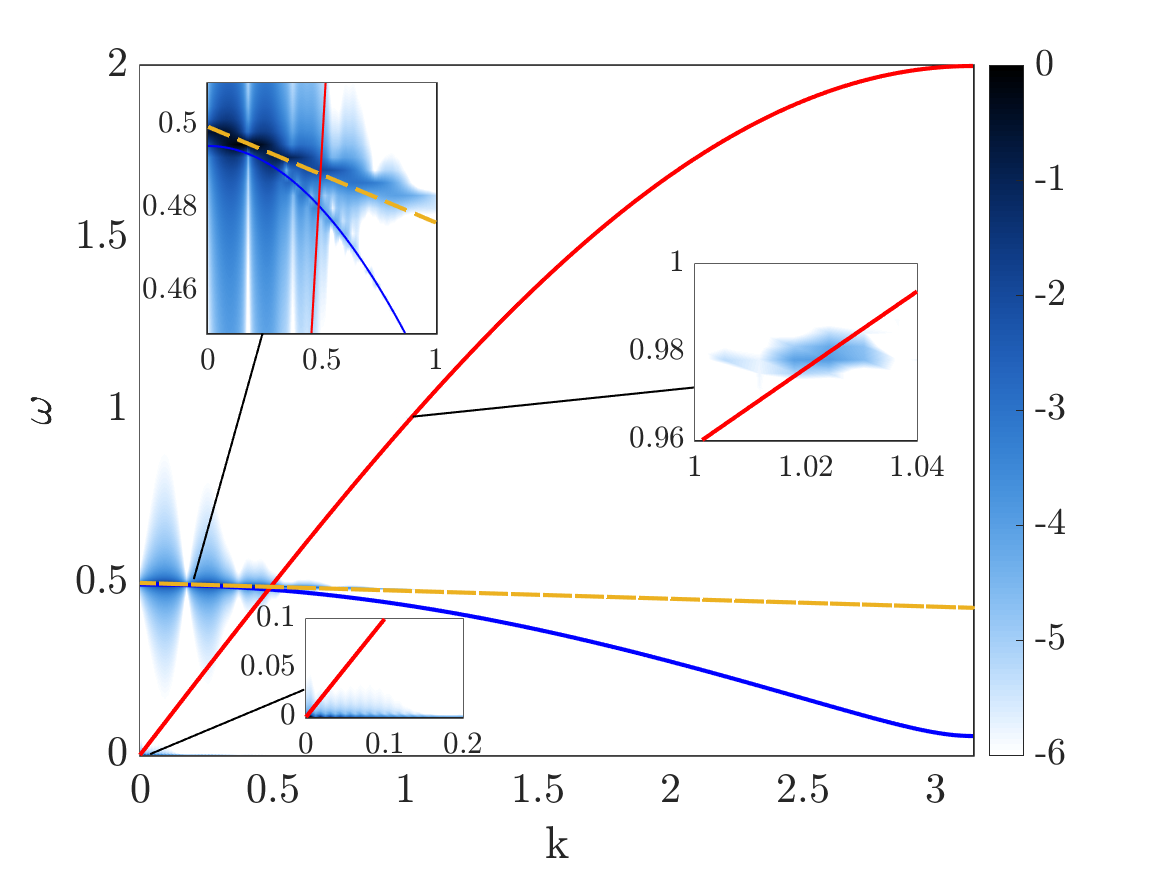}
\caption{\label{DR_white_Lego_zoom} Numerical representation of the nonlinear dispersion relation of FlexMM 2 from its dynamics, represented in Fig.\, \ref{fig:Bright_3D_Lego} using a normalized sum of the 2D-FFTs of the $\theta$ and $U$ components. The red and blue curves denote the linear dispersion relation (see Fig.\,\ref{Dispersion_relation with region stable/unstable}(b)), while the yellow line denotes NDR of the soliton described in Eq.~(\ref{NDR_soliton}). The color bar represents the $\tilde\psi(\omega,k)$, in $log$ scale.}
    \end{figure}

Following the same analysis as before, we now use as an initial condition Eqs.\,(\ref{IC_T}-\ref{IC_U}), that corresponds to a BEVS with $k = 0.1885$. The dynamics is shown in Fig. \ref{fig:Bright_3D_Lego}
The rotational dynamics is observed in panel (a), where it becomes evident that an envelope wave, in the form of a bright soliton, is propagating, accompanied by the generation of additional small waves, with frequencies around $\Omega_s$, that are radiated by the envelope wave. Note that we have confirmed that the strength of this radiation field is of the same order as that of FlexMM 1. The latter is not visible for the scales used  in Fig.~\ref{fig:Bright_3D}.
From the dynamics of the longitudinal displacement, see panel (b), we observe a kink profile followed by small waves, at frequency calculated to be around $2\Omega_s$, that move away at a relatively higher speed than the radiation observed in the rotation field.
For a better understanding of this dynamics, we examine the temporal and spatial frequency spectra in Fig.\,\ref{DR_white_Lego_zoom}.

Here, we first note that  
a large amount of the 2D FFT is centered around the line that is tangent to $\omega^{(2)}$ at point $(\Omega_s, k_s )$. This corresponds 
to the NDR of a bright soliton. We also note that again we have spectral contribution around $k=0$, coming from the dc-term, cf.~Eq.~(\ref{NDR_soliton}). These two observations are a signature of the
BEVS propagation through the lattice. However, in this case we also observe significant components of the 2D FFT in other regimes of the $\omega-k$ space. In particular, we observe frequencies around $2\Omega_s$ belonging to the dispersion of the $U$ DOF i.e., $\omega^{(1)}$ (see rightmost inset of Fig.\,\ref{DR_white_Lego_zoom}) as expected by the quadratic terms in Eq.~(\ref{Motion_equation_Order_3_NL}a). 
These frequency components corresponding to the radiation field of $U$.
Note that this was not the case for FlexMM 1, since this frequency was in the gap of the corresponding dispersion relation. 

\section{Dark envelope vector solitons
}
\label{Dark_Soliton}
\subsection{Theoretical prediction}
In the previous section, we studied the existence and propagation of BEVS solutions in the dispersion relation region corresponding to the focusing eNLS equation,  $PQ>0$. Let us now  turn our attention to the regions associated with a defocusing eNLS, where $PQ<0$. It is established that the defocusing eNLS admits the following dark soliton solution \cite{kivshar_optical_2003},

\begin{equation}
  F_1(\xi_1,\tau_2) = A_0 \text{tanh} \left[\frac{1}{L_e}(\xi_1-\xi_0)\right]e^{-iQA_0^2\tau_2} \, ,
  \label{F_1}
\end{equation}
where $A_0$ is its amplitude and $L_e$ is its width (see Eq.~(\ref{Le})).
Following the steps presented in section \ref{Bright_soliton}, we derive the subsequent analytical solution for the rotation,
\begin{equation}
    \theta_1 (X,T) =2A_0\text{tanh}\left[\frac{\epsilon}{L_e}(X-X_0- v_g T)\right]\cos[kX-\Omega T]\, ,
    \label{Theta_1_dark}
\end{equation}
where $\Omega$ is the angular frequency. Its expression is given by,
\begin{equation}
    \Omega=\omega^{(2)}+\epsilon^2 A_0^2Q \,.
\end{equation}
As for the bright soliton solution cf.~ Eqs.~(\ref{Analy_theta}-\ref{capital_omega}), the angular frequency has undergone a shift at order  $\epsilon^2$, relative to the linear $\omega^{(2)}$. This shift can manifest itself either above or below the linear branch, depending on the sign of $Q$.
The combination of Eq.~(\ref{G0}) with the envelope part of Eq.~(\ref{Theta_1_dark}) gives an expression for $G_0$ yielding,
\begin{equation}
\begin{split}
        U_0(X,T) &=\frac{A_0^2}{v_g^2-1} \left(\epsilon (X-X_0-v_gT)\right.\\
        &\left.-L_e\tanh\left[\frac{\epsilon}{L_e}(X-X_0-v_g T)\right]\right)\, .
        \end{split}
        \label{Analy_U_dark}
\end{equation}
Equations (\ref{Theta_1_dark}-\ref{Analy_U_dark}) form a polarized nonlinear wave solution 
which from now on will be called dark envelope vector soliton (DEVS). 

\subsection{Dark envelope vector soliton propagation in FlexMM}

To validate our predictions regarding the existence of DEVS, we solve the discrete set of equations (\ref{Full_discrete}) using the process described in Sec. \ref{part:BEVS}. We apply free boundary conditions at both ends of the structure and perform the integration over a duration of five nonlinear times: $t_f=5 T_{NL}$, cf.~Eq.~(\ref{NL_length}).

\begin{figure}[ht!]
\includegraphics[width=0.48\textwidth]{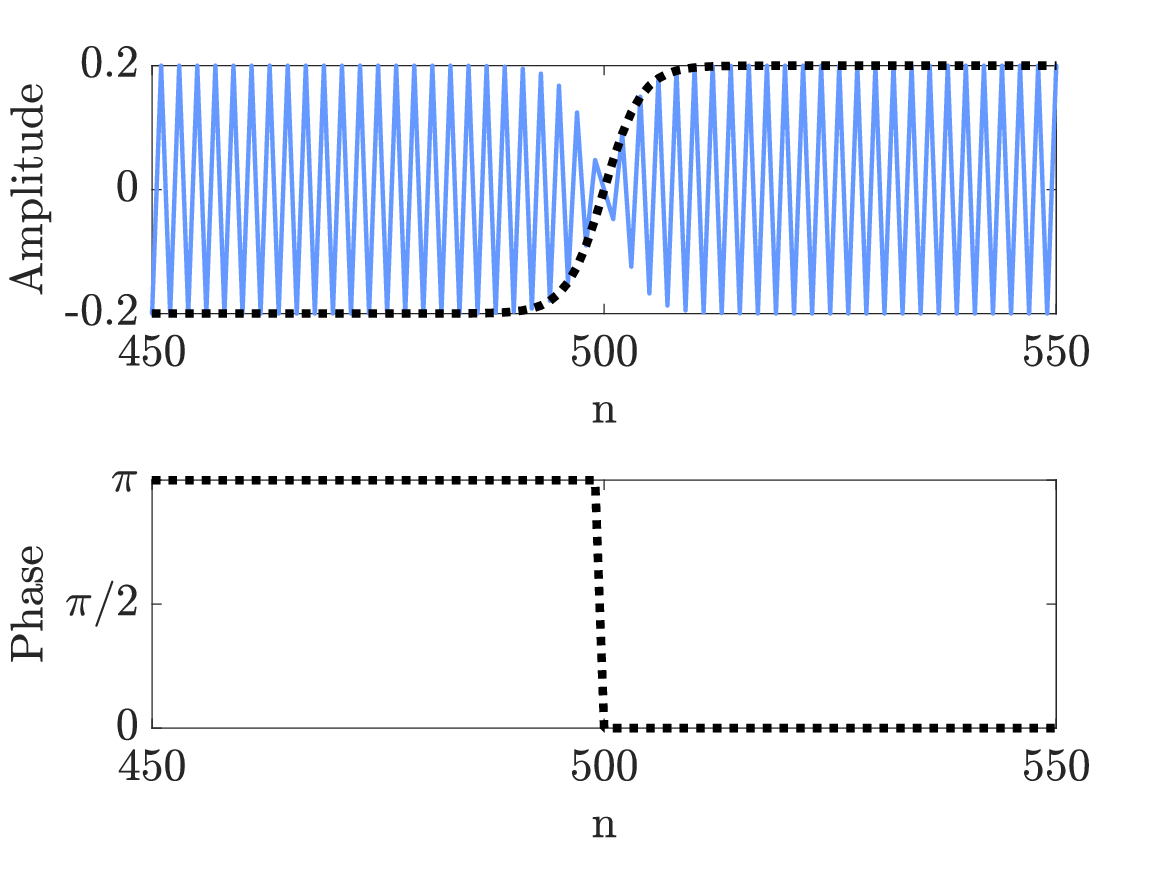}
\caption{\label{IC_dark} Analytical solution of the rotational component $\epsilon \theta_1$ (cf.~ Eq.~(\ref{Theta_1_dark})), and its corresponding phase at $T=0$, for the DEVS of wave number $k=\pi$.}
\end{figure}

In the case of DEVS, the presence of a jump in the phase field, see Fig.\,\ref{IC_dark}, leads to a mismatch with the free boundary conditions, causing boundary effects that propagate through the lattice.  
To avoid these effects, similarly to \cite{frantzeskakis_dark_2010,zaera_propagation_2018}, we multiply the $\theta$ field of Eq.~(\ref{Theta_1_dark}) by a super-gaussian window $\mathcal{W}$ of the following form,
\begin{equation}
    \mathcal{W} =e^{- \left(\frac{\xi_1-\xi_0}{S}\right)^p} =e^{- \left(\frac{(X-X_0-v_g T)}{s}\right)^p} \, ,
    \label{window}
\end{equation}
centered on the initial position $X_0$, of the dark soliton. The width of the window is governed by $s=S/ \epsilon$ where the $p$  parameter controls the edges sharpness. For the numerical simulations, we use $s= N/10=100$ and $p= 8$. The application of the spatial window $\mathcal{W}$  modulates to zero the initial rotation displacement and velocity near the boundaries. The dependence of $G_0$ on $F_1$ established by the combination of Eqs.~\ref{F1} and \ref{G0} written,
\begin{equation}
    G_0(\xi_1) =  \frac{1}{v_g^2-1}\int|F_1|^2 d\xi_1,
    \label{G_0_ini}
\end{equation}
involves that in the presence of the window $\mathcal{W}$, $G_0$ is now dependent on the product $F_1 \mathcal{W}$. Substituting $F_1$ by $F_1\mathcal{W}$ by putting Eqs.\,(\ref{F_1}-\ref{window}) in Eq.~(\ref{G_0_ini}) leads to,
\begin{equation}
\begin{split}
    G_0(\xi_1) =  \frac{A_0^2}{v_g^2-1}\int \tanh^2\left(\frac{\xi_1-\xi_0}{L_e}\right)e^{-2 \left(\frac{\xi_1-\xi_0}{S}\right)^{p}}\ d\xi_1\,.
    \end{split}
    \label{G_0_int_ind}
\end{equation}
The resulting integral has no analytical solution. Thus, the integration is numerically solved by computing the approximate cumulative integral of Eq.~(\ref{G_0_int_ind}) via the trapezoidal method for each time step of the integration. At $T=0$ the result gives the initial condition Eq.~(\ref{G_0_ini_2}). Performing the derivative on real time ($T$), noted $\dot{(\;)}$, before the numerical integration allows to obtain the initial condition for the velocity Eq.~(\ref{G_0_ini_3}).  Finally, the initial conditions employed to excite the lattice are,

\begin{subequations}
\begin{align}
        \theta(X,0) &= \epsilon\theta_1(X,0)\mathcal{W}(X,0)\, ,\\ 
    \dot{\theta}(X,0) &= \epsilon\dot{\theta_1}(X,0)\mathcal{W}(X,0)+... \, ,\label{T_1_ini} \\
    \nonumber\\
    U(\xi_1,0) &= \frac{\epsilon}{v_g^2-1}\int|F_1|^2 \mathcal{W}^2 d\xi_1
    \label{G_0_ini_2}\, ,\\ 
    \dot{U}(\xi_1,0) &= \frac{\epsilon}{v_g^2-1}\int\dot{|F_1|^2}\mathcal{W}^2 d\xi_1+ ... \,.
    \label{G_0_ini_3}
\end{align}
\label{ICs_dark}
\end{subequations}
The $\dot{\mathcal{W}}$ terms in Eqs.\,(\ref{T_1_ini} - \ref{G_0_ini_3}) can be neglected because they are proportional to $\epsilon^{p}$. Moreover in the numerical simulations $\dot{\mathcal{W}}$ is proportional to $\epsilon^{8}$.


\begin{figure}[ht!]
\includegraphics[width=0.4\textwidth]{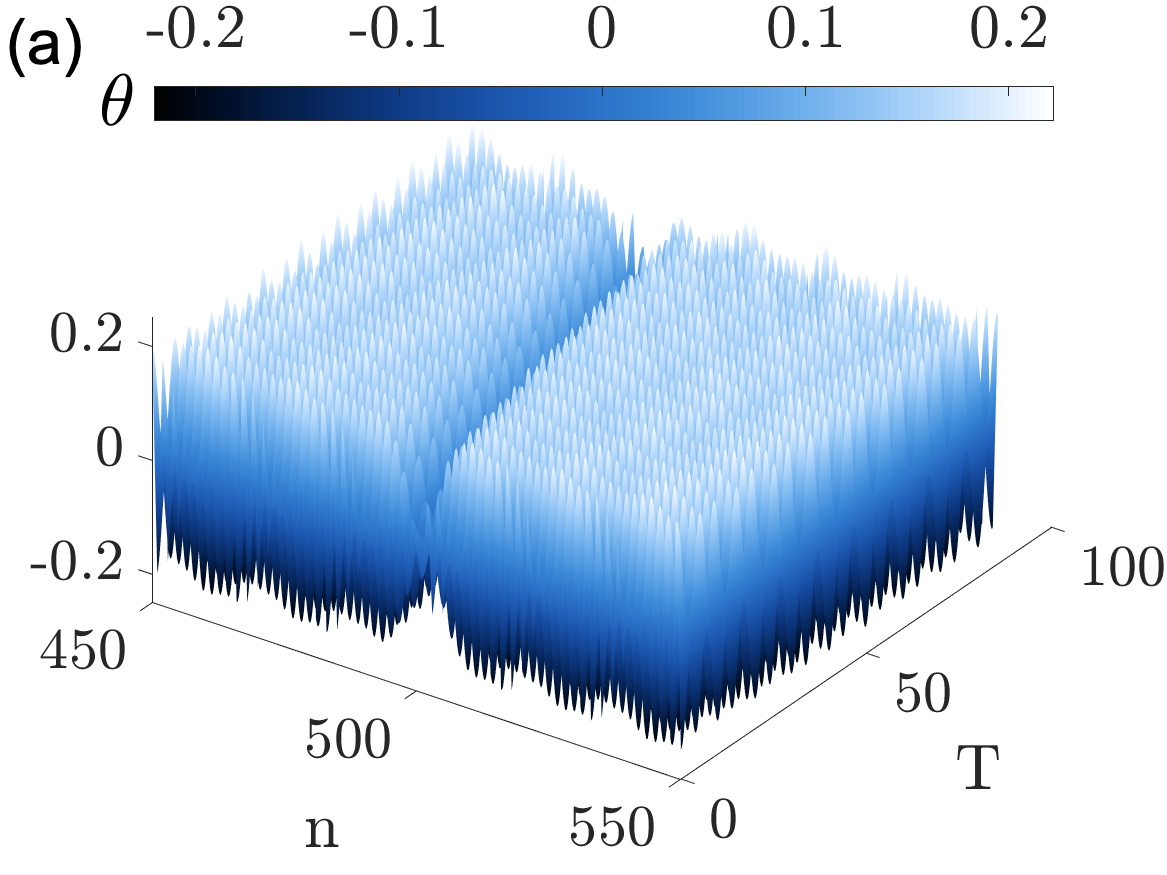}
\includegraphics[width=0.4\textwidth]{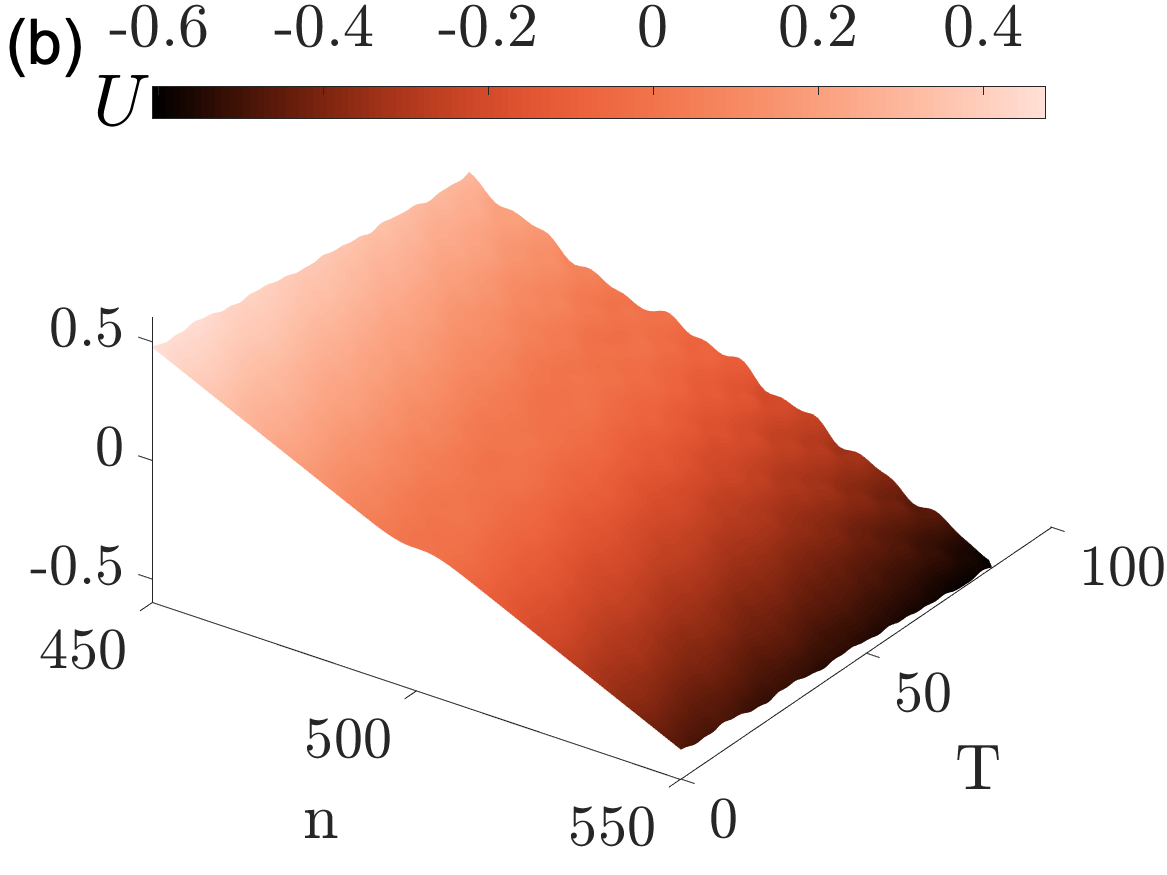}
\caption{\label{Dark_471} Panels (a-b) represent the evolution in time ($T$) of the rotational (a) and longitudinal (b) displacements along the chain ($n$). The results correspond to a FlexMM (FlexMM 1) defined by the following set of parameters : $\alpha =1.815$, $K_s =0.1851$, $K_{\theta} =1.534\text{e}^{-2}$. A DEVS is generated by the initial conditions expressed in Eqs.\,(\ref{IC_T}-\ref{IC_U}), for a spatial frequency of $k=2.9531$, and an amplitude of $\epsilon A_0=0.1$.}
\end{figure}

\begin{figure}[ht!]
\includegraphics[width=0.48\textwidth]{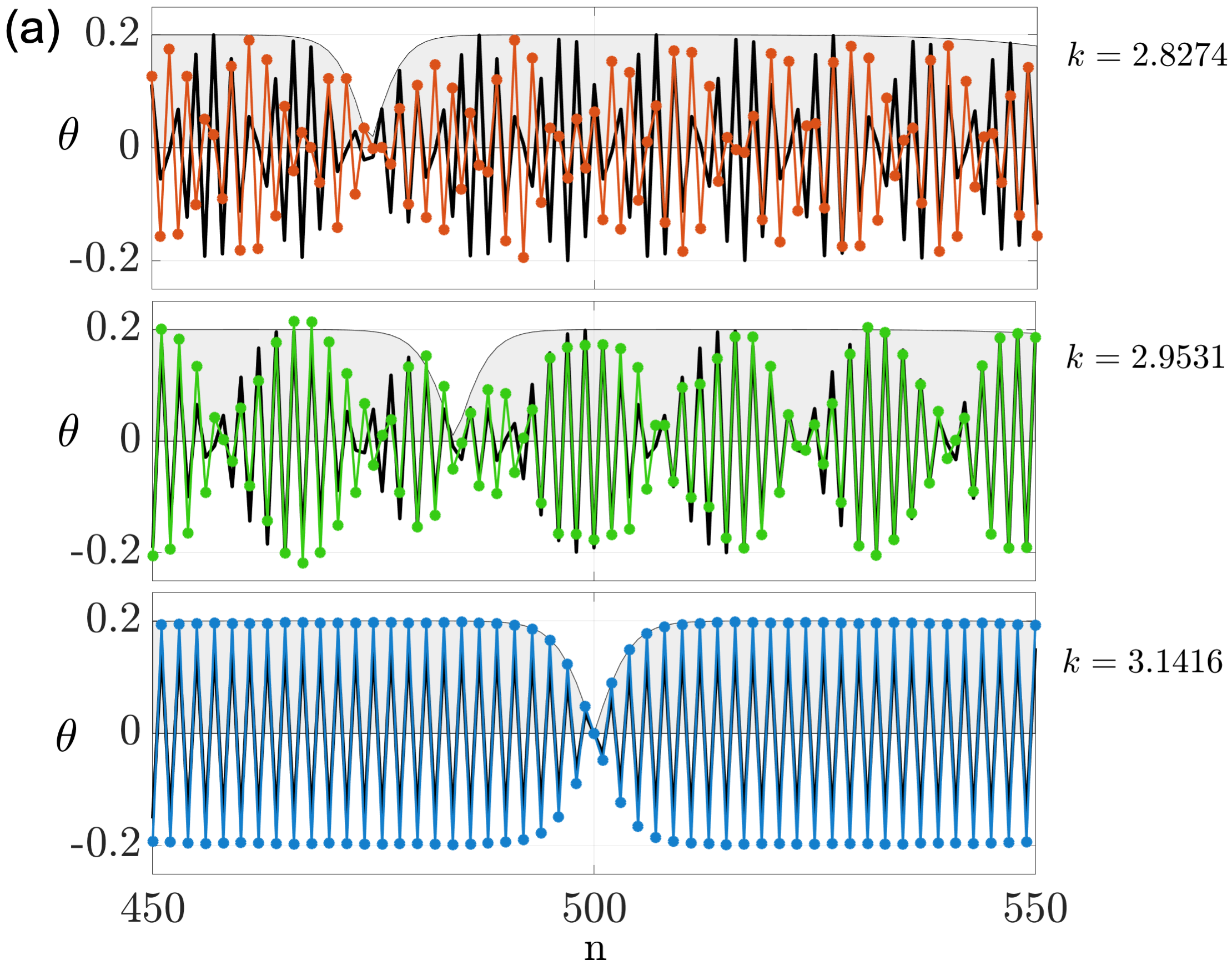}
\includegraphics[width=0.48\textwidth]{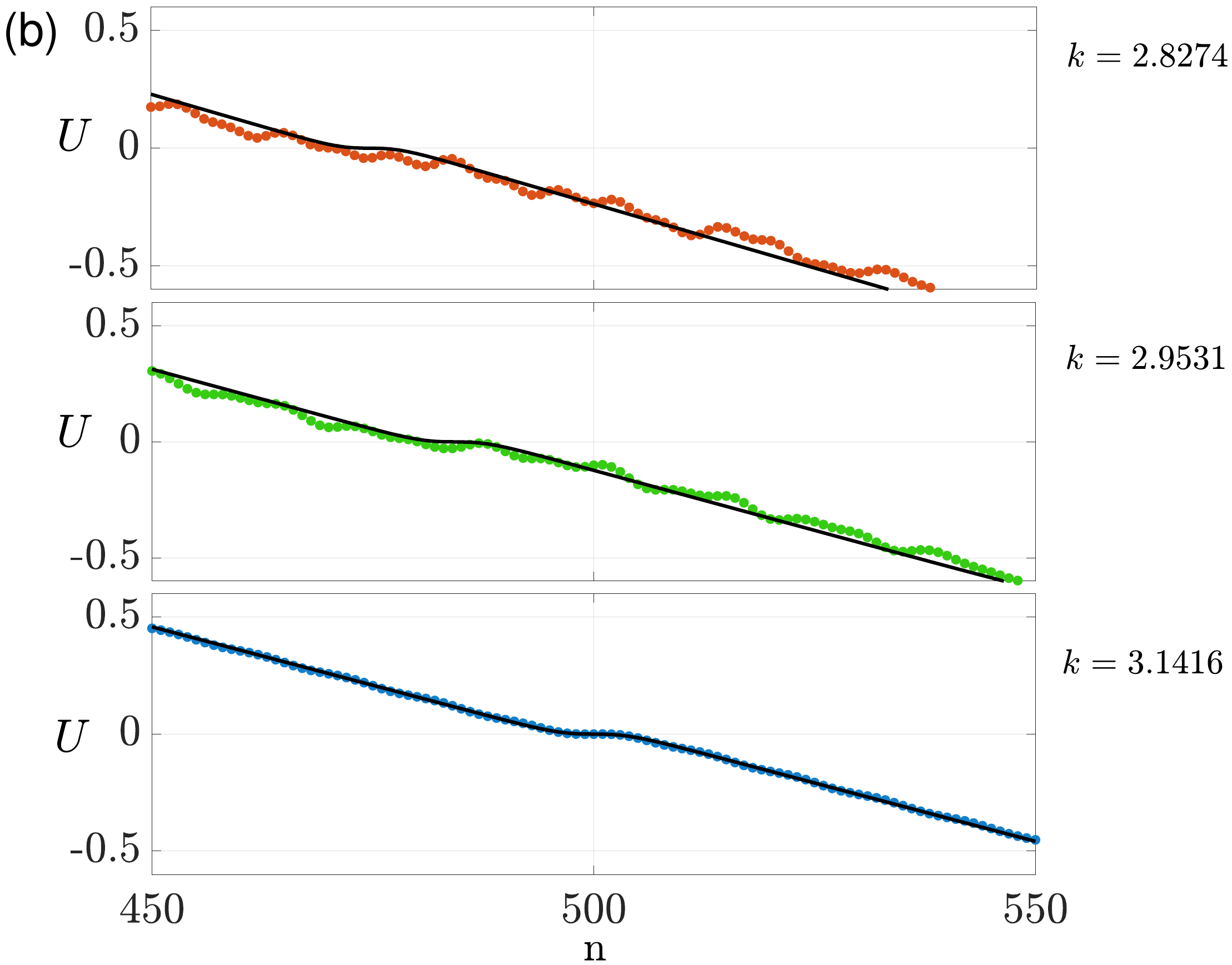}
\caption{\label{Dark_lines} Rotational and longitudinal displacement amplitudes along the chain ($n$) at final time $t_f =5 T_{NL}$. The DEVS are generated by initial conditions : a spatial frequency of $k=2.8274$ in orange , $k=2.9531$ in green, and $k=\pi$ in blue. For the three cases an amplitude of $A_0=10$ and a perturbation of $\epsilon = 0.01$ are used.}
\end{figure}

In Fig.\,\ref{Dark_471}, we show the nonlinear dynamics of the FlexMM 1, cf.~Fig.\,\ref{Dispersion_relation with region stable/unstable}(a) using as an initial condition a DEVS (Eq.~\ref{ICs_dark}) with 
$k = 2.9531$ (see green triangle in Fig.\,\ref{Dispersion_relation with region stable/unstable}(a)). As one can see in Fig. \ref{Dark_471}(a), the envelope of the rotational DOF is a continuous dip that propagates at a constant velocity and maintains its shape. Moreover, the profile of $U$, displayed in Fig.\,\ref{Dark_471}(b), also remains approximately constant in time, which is a characteristic of vector solitary waves according to our theoretical predictions. To complete the analysis, Fig.\,\ref{Dark_lines} represents DEVS at final time of integration in FlexMM 1 for initial conditions with different wave numbers indicated by colored triangles in Fig.\,\ref{Dispersion_relation with region stable/unstable}(a). We can see that the numerical results (color dotted lines) remain close to the theoretical ones, in terms of the carrier wave (black line) and of the absolute value of the envelope (gray area).  

\section{Conclusion}
In our study, FlexMMs are mechanical structures with special properties, such as geometric nonlinearity resulting from the rotation of their constituent particles. The presence of multiple DOFs and the remarkable tunability of their components provide an ideal experimental platform for the study of various nonlinear wave phenomena. In this work, we demonstrated the generation and propagation of nonlinear envelope waves under the form of bright and dark envelope vector solitons. In particular, we found that the rotational DOF can be described by an eNLS equation, and that longitudinal displacements follow the dynamics induced by the nonlinear coupling  through a dc-term at the leading order. This dc-term was not observed in our studies on modulation instability phenomena \cite{demiquel_modulation_2023}. Both analytical and numerical results show that with an appropriate choice of physical parameters for the FlexMM, in particular specific combinations of inertia $\alpha$ and stiffness parameters ($K_s$ and $K_\theta$), the propagation of these solutions in FlexMMs becomes feasible and robust. This demonstrates the significant versatility offered by the proposed system in the manipulation of weakly nonlinear modulated waves. 

 The present work demonstrates the great potential that nonlinear FlexMMs have for the observation and control of both typical and novel nonlinear phenomena related to modulated waves. Several natural extensions of this work include the excitation of the system using driving functions on one of the extremity of the chain, as well as adding dissipation in the system using for example linear viscous damping terms in the discrete set of equations of motion. Both aspects are currently under investigation and results will be presented in future publications with experimental results. Other interesting perspectives are the generation and dynamics of coherent structures such as the Peregrine soliton \cite{tikan_universality_2017}, breathers or extreme wave \cite{charalampidis_phononic_2018,tikan_local_2021} effects in nonlinear FlexMMs.
\section*{Acknowledgement}
The authors acknowledge the support from the project ExFLEM ANR-21-CE30-0003-01.

\newpage
\renewcommand\bibname{References}
\let\itshape\upshape
\normalem

\begin{thebibliography}{46}%
\makeatletter
\providecommand \@ifxundefined [1]{%
 \@ifx{#1\undefined}
}%
\providecommand \@ifnum [1]{%
 \ifnum #1\expandafter \@firstoftwo
 \else \expandafter \@secondoftwo
 \fi
}%
\providecommand \@ifx [1]{%
 \ifx #1\expandafter \@firstoftwo
 \else \expandafter \@secondoftwo
 \fi
}%
\providecommand \natexlab [1]{#1}%
\providecommand \enquote  [1]{``#1''}%
\providecommand \bibnamefont  [1]{#1}%
\providecommand \bibfnamefont [1]{#1}%
\providecommand \citenamefont [1]{#1}%
\providecommand \href@noop [0]{\@secondoftwo}%
\providecommand \href [0]{\begingroup \@sanitize@url \@href}%
\providecommand \@href[1]{\@@startlink{#1}\@@href}%
\providecommand \@@href[1]{\endgroup#1\@@endlink}%
\providecommand \@sanitize@url [0]{\catcode `\\12\catcode `\$12\catcode
  `\&12\catcode `\#12\catcode `\^12\catcode `\_12\catcode `\%12\relax}%
\providecommand \@@startlink[1]{}%
\providecommand \@@endlink[0]{}%
\providecommand \url  [0]{\begingroup\@sanitize@url \@url }%
\providecommand \@url [1]{\endgroup\@href {#1}{\urlprefix }}%
\providecommand \urlprefix  [0]{URL }%
\providecommand \Eprint [0]{\href }%
\providecommand \doibase [0]{https://doi.org/}%
\providecommand \selectlanguage [0]{\@gobble}%
\providecommand \bibinfo  [0]{\@secondoftwo}%
\providecommand \bibfield  [0]{\@secondoftwo}%
\providecommand \translation [1]{[#1]}%
\providecommand \BibitemOpen [0]{}%
\providecommand \bibitemStop [0]{}%
\providecommand \bibitemNoStop [0]{.\EOS\space}%
\providecommand \EOS [0]{\spacefactor3000\relax}%
\providecommand \BibitemShut  [1]{\csname bibitem#1\endcsname}%
\let\auto@bib@innerbib\@empty
\bibitem [{\citenamefont {Bertoldi}\ \emph {et~al.}(2017)\citenamefont
  {Bertoldi}, \citenamefont {Vitelli}, \citenamefont {Christensen},\ and\
  \citenamefont {van Hecke}}]{bertoldi_flexible_2017}%
  \BibitemOpen
  \bibfield  {author} {\bibinfo {author} {\bibfnamefont {K.}~\bibnamefont
  {Bertoldi}}, \bibinfo {author} {\bibfnamefont {V.}~\bibnamefont {Vitelli}},
  \bibinfo {author} {\bibfnamefont {J.}~\bibnamefont {Christensen}},\ and\
  \bibinfo {author} {\bibfnamefont {M.}~\bibnamefont {van Hecke}},\ }\bibfield
  {title} {{\selectlanguage {english}\bibinfo {title} {Flexible mechanical
  metamaterials}},\ }\href {https://doi.org/10.1038/natrevmats.2017.66}
  {\bibfield  {journal} {\bibinfo  {journal} {Nat Rev Mater}\ }\textbf
  {\bibinfo {volume} {2}},\ \bibinfo {pages} {17066} (\bibinfo {year}
  {2017})}\BibitemShut {NoStop}%
\bibitem [{\citenamefont {Filipov}\ \emph {et~al.}(2015)\citenamefont
  {Filipov}, \citenamefont {Tachi},\ and\ \citenamefont
  {Paulino}}]{filipov_origami_2015}%
  \BibitemOpen
  \bibfield  {author} {\bibinfo {author} {\bibfnamefont {E.~T.}\ \bibnamefont
  {Filipov}}, \bibinfo {author} {\bibfnamefont {T.}~\bibnamefont {Tachi}},\
  and\ \bibinfo {author} {\bibfnamefont {G.~H.}\ \bibnamefont {Paulino}},\
  }\bibfield  {title} {{\selectlanguage {english}\bibinfo {title} {Origami
  tubes assembled into stiff, yet reconfigurable structures and
  metamaterials}},\ }\href {https://doi.org/10.1073/pnas.1509465112} {\bibfield
   {journal} {\bibinfo  {journal} {Proc. Natl. Acad. Sci. U.S.A.}\ }\textbf
  {\bibinfo {volume} {112}},\ \bibinfo {pages} {12321} (\bibinfo {year}
  {2015})}\BibitemShut {NoStop}%
\bibitem [{\citenamefont {Miyazawa}\ \emph {et~al.}(2022)\citenamefont
  {Miyazawa}, \citenamefont {Chen}, \citenamefont {Chaunsali}, \citenamefont
  {Gormley}, \citenamefont {Yin}, \citenamefont {Theocharis},\ and\
  \citenamefont {Yang}}]{miyazawa_topological_2022}%
  \BibitemOpen
  \bibfield  {author} {\bibinfo {author} {\bibfnamefont {Y.}~\bibnamefont
  {Miyazawa}}, \bibinfo {author} {\bibfnamefont {C.-W.}\ \bibnamefont {Chen}},
  \bibinfo {author} {\bibfnamefont {R.}~\bibnamefont {Chaunsali}}, \bibinfo
  {author} {\bibfnamefont {T.~S.}\ \bibnamefont {Gormley}}, \bibinfo {author}
  {\bibfnamefont {G.}~\bibnamefont {Yin}}, \bibinfo {author} {\bibfnamefont
  {G.}~\bibnamefont {Theocharis}},\ and\ \bibinfo {author} {\bibfnamefont
  {J.}~\bibnamefont {Yang}},\ }\bibfield  {title} {{\selectlanguage
  {english}\bibinfo {title} {Topological state transfer in {Kresling}
  origami}},\ }\href {https://doi.org/10.1038/s43246-022-00280-0} {\bibfield
  {journal} {\bibinfo  {journal} {Commun Mater}\ }\textbf {\bibinfo {volume}
  {3}},\ \bibinfo {pages} {62} (\bibinfo {year} {2022})}\BibitemShut {NoStop}%
\bibitem [{\citenamefont {Shyu}\ \emph {et~al.}(2015)\citenamefont {Shyu},
  \citenamefont {Damasceno}, \citenamefont {Dodd}, \citenamefont {Lamoureux},
  \citenamefont {Xu}, \citenamefont {Shlian}, \citenamefont {Shtein},
  \citenamefont {Glotzer},\ and\ \citenamefont {Kotov}}]{shyu_kirigami_2015}%
  \BibitemOpen
  \bibfield  {author} {\bibinfo {author} {\bibfnamefont {T.~C.}\ \bibnamefont
  {Shyu}}, \bibinfo {author} {\bibfnamefont {P.~F.}\ \bibnamefont {Damasceno}},
  \bibinfo {author} {\bibfnamefont {P.~M.}\ \bibnamefont {Dodd}}, \bibinfo
  {author} {\bibfnamefont {A.}~\bibnamefont {Lamoureux}}, \bibinfo {author}
  {\bibfnamefont {L.}~\bibnamefont {Xu}}, \bibinfo {author} {\bibfnamefont
  {M.}~\bibnamefont {Shlian}}, \bibinfo {author} {\bibfnamefont
  {M.}~\bibnamefont {Shtein}}, \bibinfo {author} {\bibfnamefont {S.~C.}\
  \bibnamefont {Glotzer}},\ and\ \bibinfo {author} {\bibfnamefont {N.~A.}\
  \bibnamefont {Kotov}},\ }\bibfield  {title} {{\selectlanguage
  {english}\bibinfo {title} {A kirigami approach to engineering elasticity in
  nanocomposites through patterned defects}},\ }\href
  {https://doi.org/10.1038/nmat4327} {\bibfield  {journal} {\bibinfo  {journal}
  {Nature Mater}\ }\textbf {\bibinfo {volume} {14}},\ \bibinfo {pages} {785}
  (\bibinfo {year} {2015})}\BibitemShut {NoStop}%
\bibitem [{\citenamefont {Isobe}\ and\ \citenamefont
  {Okumura}(2016)}]{isobe_initial_2016}%
  \BibitemOpen
  \bibfield  {author} {\bibinfo {author} {\bibfnamefont {M.}~\bibnamefont
  {Isobe}}\ and\ \bibinfo {author} {\bibfnamefont {K.}~\bibnamefont
  {Okumura}},\ }\bibfield  {title} {{\selectlanguage {english}\bibinfo {title}
  {Initial rigid response and softening transition of highly stretchable
  kirigami sheet materials}},\ }\href {https://doi.org/10.1038/srep24758}
  {\bibfield  {journal} {\bibinfo  {journal} {Sci Rep}\ }\textbf {\bibinfo
  {volume} {6}},\ \bibinfo {pages} {24758} (\bibinfo {year}
  {2016})}\BibitemShut {NoStop}%
\bibitem [{\citenamefont {Raney}\ and\ \citenamefont
  {Lewis}(2015)}]{raney_printing_2015}%
  \BibitemOpen
  \bibfield  {author} {\bibinfo {author} {\bibfnamefont {J.~R.}\ \bibnamefont
  {Raney}}\ and\ \bibinfo {author} {\bibfnamefont {J.~A.}\ \bibnamefont
  {Lewis}},\ }\bibfield  {title} {{\selectlanguage {english}\bibinfo {title}
  {Printing mesoscale architectures}},\ }\href
  {https://doi.org/10.1557/mrs.2015.235} {\bibfield  {journal} {\bibinfo
  {journal} {MRS Bull.}\ }\textbf {\bibinfo {volume} {40}},\ \bibinfo {pages}
  {943} (\bibinfo {year} {2015})}\BibitemShut {NoStop}%
\bibitem [{\citenamefont {Sundaram}\ \emph {et~al.}(2019)\citenamefont
  {Sundaram}, \citenamefont {Skouras}, \citenamefont {Kim}, \citenamefont {Van
  Den~Heuvel},\ and\ \citenamefont {Matusik}}]{sundaram_topology_2019}%
  \BibitemOpen
  \bibfield  {author} {\bibinfo {author} {\bibfnamefont {S.}~\bibnamefont
  {Sundaram}}, \bibinfo {author} {\bibfnamefont {M.}~\bibnamefont {Skouras}},
  \bibinfo {author} {\bibfnamefont {D.~S.}\ \bibnamefont {Kim}}, \bibinfo
  {author} {\bibfnamefont {L.}~\bibnamefont {Van Den~Heuvel}},\ and\ \bibinfo
  {author} {\bibfnamefont {W.}~\bibnamefont {Matusik}},\ }\bibfield  {title}
  {{\selectlanguage {english}\bibinfo {title} {Topology optimization and {3D}
  printing of multimaterial magnetic actuators and displays}},\ }\href
  {https://doi.org/10.1126/sciadv.aaw1160} {\bibfield  {journal} {\bibinfo
  {journal} {Sci. Adv.}\ }\textbf {\bibinfo {volume} {5}},\ \bibinfo {pages}
  {eaaw1160} (\bibinfo {year} {2019})}\BibitemShut {NoStop}%
\bibitem [{\citenamefont {Deng}\ \emph {et~al.}(2021)\citenamefont {Deng},
  \citenamefont {Raney}, \citenamefont {Bertoldi},\ and\ \citenamefont
  {Tournat}}]{deng_nonlinear_2021}%
  \BibitemOpen
  \bibfield  {author} {\bibinfo {author} {\bibfnamefont {B.}~\bibnamefont
  {Deng}}, \bibinfo {author} {\bibfnamefont {J.~R.}\ \bibnamefont {Raney}},
  \bibinfo {author} {\bibfnamefont {K.}~\bibnamefont {Bertoldi}},\ and\
  \bibinfo {author} {\bibfnamefont {V.}~\bibnamefont {Tournat}},\ }\bibfield
  {title} {{\selectlanguage {english}\bibinfo {title} {Nonlinear waves in
  flexible mechanical metamaterials}},\ }\href
  {https://doi.org/10.1063/5.0050271} {\bibfield  {journal} {\bibinfo
  {journal} {Journal of Applied Physics}\ }\textbf {\bibinfo {volume} {130}},\
  \bibinfo {pages} {040901} (\bibinfo {year} {2021})}\BibitemShut {NoStop}%
\bibitem [{\citenamefont {Deng}\ \emph {et~al.}(2017)\citenamefont {Deng},
  \citenamefont {Raney}, \citenamefont {Tournat},\ and\ \citenamefont
  {Bertoldi}}]{deng_elastic_2017}%
  \BibitemOpen
  \bibfield  {author} {\bibinfo {author} {\bibfnamefont {B.}~\bibnamefont
  {Deng}}, \bibinfo {author} {\bibfnamefont {J.~R.}\ \bibnamefont {Raney}},
  \bibinfo {author} {\bibfnamefont {V.}~\bibnamefont {Tournat}},\ and\ \bibinfo
  {author} {\bibfnamefont {K.}~\bibnamefont {Bertoldi}},\ }\bibfield  {title}
  {{\selectlanguage {english}\bibinfo {title} {Elastic {Vector} {Solitons} in
  {Soft} {Architected} {Materials}}},\ }\href
  {https://doi.org/10.1103/PhysRevLett.118.204102} {\bibfield  {journal}
  {\bibinfo  {journal} {Phys. Rev. Lett.}\ }\textbf {\bibinfo {volume} {118}},\
  \bibinfo {pages} {204102} (\bibinfo {year} {2017})}\BibitemShut {NoStop}%
\bibitem [{\citenamefont {Deng}\ \emph {et~al.}(2018)\citenamefont {Deng},
  \citenamefont {Wang}, \citenamefont {He}, \citenamefont {Tournat},\ and\
  \citenamefont {Bertoldi}}]{deng_metamaterials_2018}%
  \BibitemOpen
  \bibfield  {author} {\bibinfo {author} {\bibfnamefont {B.}~\bibnamefont
  {Deng}}, \bibinfo {author} {\bibfnamefont {P.}~\bibnamefont {Wang}}, \bibinfo
  {author} {\bibfnamefont {Q.}~\bibnamefont {He}}, \bibinfo {author}
  {\bibfnamefont {V.}~\bibnamefont {Tournat}},\ and\ \bibinfo {author}
  {\bibfnamefont {K.}~\bibnamefont {Bertoldi}},\ }\bibfield  {title}
  {{\selectlanguage {english}\bibinfo {title} {Metamaterials with amplitude
  gaps for elastic solitons}},\ }\href
  {https://doi.org/10.1038/s41467-018-05908-9} {\bibfield  {journal} {\bibinfo
  {journal} {Nat Commun}\ }\textbf {\bibinfo {volume} {9}},\ \bibinfo {pages}
  {3410} (\bibinfo {year} {2018})}\BibitemShut {NoStop}%
\bibitem [{\citenamefont {Deng}\ \emph
  {et~al.}(2019{\natexlab{a}})\citenamefont {Deng}, \citenamefont {Tournat},
  \citenamefont {Wang},\ and\ \citenamefont {Bertoldi}}]{deng_anomalous_2019}%
  \BibitemOpen
  \bibfield  {author} {\bibinfo {author} {\bibfnamefont {B.}~\bibnamefont
  {Deng}}, \bibinfo {author} {\bibfnamefont {V.}~\bibnamefont {Tournat}},
  \bibinfo {author} {\bibfnamefont {P.}~\bibnamefont {Wang}},\ and\ \bibinfo
  {author} {\bibfnamefont {K.}~\bibnamefont {Bertoldi}},\ }\bibfield  {title}
  {{\selectlanguage {english}\bibinfo {title} {Anomalous {Collisions} of
  {Elastic} {Vector} {Solitons} in {Mechanical} {Metamaterials}}},\ }\href
  {https://doi.org/10.1103/PhysRevLett.122.044101} {\bibfield  {journal}
  {\bibinfo  {journal} {Phys. Rev. Lett.}\ }\textbf {\bibinfo {volume} {122}},\
  \bibinfo {pages} {044101} (\bibinfo {year} {2019}{\natexlab{a}})}\BibitemShut
  {NoStop}%
\bibitem [{\citenamefont {Herbold}\ and\ \citenamefont
  {Nesterenko}(2013)}]{herbold_propagation_2013}%
  \BibitemOpen
  \bibfield  {author} {\bibinfo {author} {\bibfnamefont {E.~B.}\ \bibnamefont
  {Herbold}}\ and\ \bibinfo {author} {\bibfnamefont {V.~F.}\ \bibnamefont
  {Nesterenko}},\ }\bibfield  {title} {{\selectlanguage {english}\bibinfo
  {title} {Propagation of {Rarefaction} {Pulses} in {Discrete} {Materials} with
  {Strain}-{Softening} {Behavior}}},\ }\href
  {https://doi.org/10.1103/PhysRevLett.110.144101} {\bibfield  {journal}
  {\bibinfo  {journal} {Phys. Rev. Lett.}\ }\textbf {\bibinfo {volume} {110}},\
  \bibinfo {pages} {144101} (\bibinfo {year} {2013})}\BibitemShut {NoStop}%
\bibitem [{\citenamefont {Deng}\ \emph
  {et~al.}(2019{\natexlab{b}})\citenamefont {Deng}, \citenamefont {Zhang},
  \citenamefont {He}, \citenamefont {Tournat}, \citenamefont {Wang},\ and\
  \citenamefont {Bertoldi}}]{deng_propagation_2019}%
  \BibitemOpen
  \bibfield  {author} {\bibinfo {author} {\bibfnamefont {B.}~\bibnamefont
  {Deng}}, \bibinfo {author} {\bibfnamefont {Y.}~\bibnamefont {Zhang}},
  \bibinfo {author} {\bibfnamefont {Q.}~\bibnamefont {He}}, \bibinfo {author}
  {\bibfnamefont {V.}~\bibnamefont {Tournat}}, \bibinfo {author} {\bibfnamefont
  {P.}~\bibnamefont {Wang}},\ and\ \bibinfo {author} {\bibfnamefont
  {K.}~\bibnamefont {Bertoldi}},\ }\bibfield  {title} {{\selectlanguage
  {english}\bibinfo {title} {Propagation of elastic solitons in chains of
  pre-deformed beams}},\ }\href {https://doi.org/10.1088/1367-2630/ab2810}
  {\bibfield  {journal} {\bibinfo  {journal} {New J. Phys.}\ }\textbf {\bibinfo
  {volume} {21}},\ \bibinfo {pages} {073008} (\bibinfo {year}
  {2019}{\natexlab{b}})}\BibitemShut {NoStop}%
\bibitem [{\citenamefont {Jin}\ \emph {et~al.}(2020)\citenamefont {Jin},
  \citenamefont {Khajehtourian}, \citenamefont {Mueller}, \citenamefont
  {Rafsanjani}, \citenamefont {Tournat}, \citenamefont {Bertoldi},\ and\
  \citenamefont {Kochmann}}]{jin_guided_2020}%
  \BibitemOpen
  \bibfield  {author} {\bibinfo {author} {\bibfnamefont {L.}~\bibnamefont
  {Jin}}, \bibinfo {author} {\bibfnamefont {R.}~\bibnamefont {Khajehtourian}},
  \bibinfo {author} {\bibfnamefont {J.}~\bibnamefont {Mueller}}, \bibinfo
  {author} {\bibfnamefont {A.}~\bibnamefont {Rafsanjani}}, \bibinfo {author}
  {\bibfnamefont {V.}~\bibnamefont {Tournat}}, \bibinfo {author} {\bibfnamefont
  {K.}~\bibnamefont {Bertoldi}},\ and\ \bibinfo {author} {\bibfnamefont
  {D.~M.}\ \bibnamefont {Kochmann}},\ }\bibfield  {title} {{\selectlanguage
  {english}\bibinfo {title} {Guided transition waves in multistable mechanical
  metamaterials}},\ }\href {https://doi.org/10.1073/pnas.1913228117} {\bibfield
   {journal} {\bibinfo  {journal} {Proc. Natl. Acad. Sci. U.S.A.}\ }\textbf
  {\bibinfo {volume} {117}},\ \bibinfo {pages} {2319} (\bibinfo {year}
  {2020})}\BibitemShut {NoStop}%
\bibitem [{\citenamefont {Zareei}\ \emph {et~al.}(2020)\citenamefont {Zareei},
  \citenamefont {Deng},\ and\ \citenamefont
  {Bertoldi}}]{zareei_harnessing_2020}%
  \BibitemOpen
  \bibfield  {author} {\bibinfo {author} {\bibfnamefont {A.}~\bibnamefont
  {Zareei}}, \bibinfo {author} {\bibfnamefont {B.}~\bibnamefont {Deng}},\ and\
  \bibinfo {author} {\bibfnamefont {K.}~\bibnamefont {Bertoldi}},\ }\bibfield
  {title} {{\selectlanguage {english}\bibinfo {title} {Harnessing transition
  waves to realize deployable structures}},\ }\href
  {https://doi.org/10.1073/pnas.1917887117} {\bibfield  {journal} {\bibinfo
  {journal} {Proc. Natl. Acad. Sci. U.S.A.}\ }\textbf {\bibinfo {volume}
  {117}},\ \bibinfo {pages} {4015} (\bibinfo {year} {2020})}\BibitemShut
  {NoStop}%
\bibitem [{\citenamefont {Yasuda}\ \emph {et~al.}(2020)\citenamefont {Yasuda},
  \citenamefont {Korpas},\ and\ \citenamefont
  {Raney}}]{yasuda_transition_2020}%
  \BibitemOpen
  \bibfield  {author} {\bibinfo {author} {\bibfnamefont {H.}~\bibnamefont
  {Yasuda}}, \bibinfo {author} {\bibfnamefont {L.~M.}\ \bibnamefont {Korpas}},\
  and\ \bibinfo {author} {\bibfnamefont {J.~R.}\ \bibnamefont {Raney}},\
  }\bibfield  {title} {{\selectlanguage {english}\bibinfo {title} {Transition
  {Waves} and {Formation} of {Domain} {Walls} in {Multistable} {Mechanical}
  {Metamaterials}}},\ }\href {https://doi.org/10.1103/PhysRevApplied.13.054067}
  {\bibfield  {journal} {\bibinfo  {journal} {Phys. Rev. Applied}\ }\textbf
  {\bibinfo {volume} {13}},\ \bibinfo {pages} {054067} (\bibinfo {year}
  {2020})}\BibitemShut {NoStop}%
\bibitem [{\citenamefont {Demiquel}\ \emph {et~al.}(2023)\citenamefont
  {Demiquel}, \citenamefont {Achilleos}, \citenamefont {Theocharis},\ and\
  \citenamefont {Tournat}}]{demiquel_modulation_2023}%
  \BibitemOpen
  \bibfield  {author} {\bibinfo {author} {\bibfnamefont {A.}~\bibnamefont
  {Demiquel}}, \bibinfo {author} {\bibfnamefont {V.}~\bibnamefont {Achilleos}},
  \bibinfo {author} {\bibfnamefont {G.}~\bibnamefont {Theocharis}},\ and\
  \bibinfo {author} {\bibfnamefont {V.}~\bibnamefont {Tournat}},\ }\bibfield
  {title} {{\selectlanguage {english}\bibinfo {title} {Modulation instability
  in nonlinear flexible mechanical metamaterials}},\ }\href
  {https://doi.org/10.1103/PhysRevE.107.054212} {\bibfield  {journal} {\bibinfo
   {journal} {Phys. Rev. E}\ }\textbf {\bibinfo {volume} {107}},\ \bibinfo
  {pages} {054212} (\bibinfo {year} {2023})}\BibitemShut {NoStop}%
\bibitem [{\citenamefont {Ablowitz}\ \emph {et~al.}(2004)\citenamefont
  {Ablowitz}, \citenamefont {Prinari},\ and\ \citenamefont
  {Trubatch}}]{ablowitz_discrete_2004}%
  \BibitemOpen
  \bibfield  {author} {\bibinfo {author} {\bibfnamefont {M.~J.}\ \bibnamefont
  {Ablowitz}}, \bibinfo {author} {\bibfnamefont {B.}~\bibnamefont {Prinari}},\
  and\ \bibinfo {author} {\bibfnamefont {A.~D.}\ \bibnamefont {Trubatch}},\
  }\href@noop {} {{\selectlanguage {english}\emph {\bibinfo {title} {Discrete
  and {Continuous} {Nonlinear} {Schrodinger} {Systems}}}}},\ \bibinfo {edition}
  {cambridge university press}\ ed.\ (\bibinfo {year} {2004})\BibitemShut
  {NoStop}%
\bibitem [{\citenamefont {Peyrard}\ and\ \citenamefont
  {Dauxois}(2010)}]{peyrard_physics_2010}%
  \BibitemOpen
  \bibfield  {author} {\bibinfo {author} {\bibfnamefont {M.}~\bibnamefont
  {Peyrard}}\ and\ \bibinfo {author} {\bibfnamefont {T.}~\bibnamefont
  {Dauxois}},\ }\bibfield  {title} {{\selectlanguage {english}\bibinfo {title}
  {Physics of solitons}},\ }in\ \href@noop {} {{\selectlanguage {english}\emph
  {\bibinfo {booktitle} {Physics of solitons}}}}\ (\bibinfo {year} {2010})\
  \bibinfo {edition} {cambridge university press}\ ed.,\ pp.\ \bibinfo {pages}
  {71--109}\BibitemShut {NoStop}%
\bibitem [{\citenamefont {Silberberg}(1990)}]{silberberg_collapse_1990}%
  \BibitemOpen
  \bibfield  {author} {\bibinfo {author} {\bibfnamefont {Y.}~\bibnamefont
  {Silberberg}},\ }\bibfield  {title} {\bibinfo {title} {Collapse of otical
  pulses},\ }\href {https://doi.org/10.1364/OL.15.001282} {\bibfield  {journal}
  {\bibinfo  {journal} {Opt. Lett.}\ }\textbf {\bibinfo {volume} {15}},\
  \bibinfo {pages} {1282} (\bibinfo {year} {1990})}\BibitemShut {NoStop}%
\bibitem [{\citenamefont {Hasegawa}\ and\ \citenamefont
  {Tappert}(1973)}]{hasegawa_transmission_1973}%
  \BibitemOpen
  \bibfield  {author} {\bibinfo {author} {\bibfnamefont {A.}~\bibnamefont
  {Hasegawa}}\ and\ \bibinfo {author} {\bibfnamefont {F.}~\bibnamefont
  {Tappert}},\ }\bibfield  {title} {\bibinfo {title} {Transmission of
  stationary nonlinear optical pulses in dispersive dielectric fibers. {I}.
  {Anomalous} dispersion},\ }\href@noop {} {\bibfield  {journal} {\bibinfo
  {journal} {Appl. Phys. Lett.}\ }\textbf {\bibinfo {volume} {23}},\ \bibinfo
  {pages} {142} (\bibinfo {year} {1973})}\BibitemShut {NoStop}%
\bibitem [{\citenamefont {Mollenauer}\ \emph {et~al.}(1980)\citenamefont
  {Mollenauer}, \citenamefont {Stolen},\ and\ \citenamefont
  {Gordon}}]{mollenauer_experimental_1980}%
  \BibitemOpen
  \bibfield  {author} {\bibinfo {author} {\bibfnamefont {L.~F.}\ \bibnamefont
  {Mollenauer}}, \bibinfo {author} {\bibfnamefont {R.~H.}\ \bibnamefont
  {Stolen}},\ and\ \bibinfo {author} {\bibfnamefont {J.~P.}\ \bibnamefont
  {Gordon}},\ }\bibfield  {title} {{\selectlanguage {english}\bibinfo {title}
  {Experimental {Observation} of {Picosecond} {Pulse} {Narrowing} and
  {Solitons} in {Optical} {Fibers}}},\ }\href
  {https://doi.org/10.1103/PhysRevLett.45.1095} {\bibfield  {journal} {\bibinfo
   {journal} {Phys. Rev. Lett.}\ }\textbf {\bibinfo {volume} {45}},\ \bibinfo
  {pages} {1095} (\bibinfo {year} {1980})}\BibitemShut {NoStop}%
\bibitem [{\citenamefont {Pitaevskii}\ and\ \citenamefont
  {Stringari}(2016)}]{pitaevskii_bose-einstein_2016}%
  \BibitemOpen
  \bibfield  {author} {\bibinfo {author} {\bibfnamefont {L.}~\bibnamefont
  {Pitaevskii}}\ and\ \bibinfo {author} {\bibfnamefont {S.}~\bibnamefont
  {Stringari}},\ }\href@noop {} {\emph {\bibinfo {title} {Bose-{Einstein}
  condensation and superfluidity}}},\ Vol.\ \bibinfo {volume} {164}\ (\bibinfo
  {publisher} {Oxford University Press.},\ \bibinfo {year} {2016})\BibitemShut
  {NoStop}%
\bibitem [{\citenamefont {Chabchoub}\ \emph {et~al.}(2013)\citenamefont
  {Chabchoub}, \citenamefont {Kimmoun}, \citenamefont {Branger}, \citenamefont
  {Hoffmann}, \citenamefont {Proment}, \citenamefont {Onorato},\ and\
  \citenamefont {Akhmediev}}]{chabchoub_experimental_2013}%
  \BibitemOpen
  \bibfield  {author} {\bibinfo {author} {\bibfnamefont {A.}~\bibnamefont
  {Chabchoub}}, \bibinfo {author} {\bibfnamefont {O.}~\bibnamefont {Kimmoun}},
  \bibinfo {author} {\bibfnamefont {H.}~\bibnamefont {Branger}}, \bibinfo
  {author} {\bibfnamefont {N.}~\bibnamefont {Hoffmann}}, \bibinfo {author}
  {\bibfnamefont {D.}~\bibnamefont {Proment}}, \bibinfo {author} {\bibfnamefont
  {M.}~\bibnamefont {Onorato}},\ and\ \bibinfo {author} {\bibfnamefont
  {N.}~\bibnamefont {Akhmediev}},\ }\bibfield  {title} {{\selectlanguage
  {english}\bibinfo {title} {Experimental {Observation} of {Dark} {Solitons} on
  the {Surface} of {Water}}},\ }\href
  {https://doi.org/10.1103/PhysRevLett.110.124101} {\bibfield  {journal}
  {\bibinfo  {journal} {Physical Review Letters}\ }\textbf {\bibinfo {volume}
  {110}},\ \bibinfo {pages} {124101} (\bibinfo {year} {2013})}\BibitemShut
  {NoStop}%
\bibitem [{\citenamefont {Emplit}\ \emph {et~al.}(1987)\citenamefont {Emplit},
  \citenamefont {Hamaide}, \citenamefont {Reynaud}, \citenamefont {Froehly},\
  and\ \citenamefont {Barthelemy}}]{emplit_picosecond_1987}%
  \BibitemOpen
  \bibfield  {author} {\bibinfo {author} {\bibfnamefont {P.}~\bibnamefont
  {Emplit}}, \bibinfo {author} {\bibfnamefont {J.}~\bibnamefont {Hamaide}},
  \bibinfo {author} {\bibfnamefont {F.}~\bibnamefont {Reynaud}}, \bibinfo
  {author} {\bibfnamefont {C.}~\bibnamefont {Froehly}},\ and\ \bibinfo {author}
  {\bibfnamefont {A.}~\bibnamefont {Barthelemy}},\ }\bibfield  {title}
  {{\selectlanguage {english}\bibinfo {title} {Picosecond steps and dark pulses
  through nonlinear single mode fibers}},\ }\href
  {https://doi.org/10.1016/0030-4018(87)90003-4} {\bibfield  {journal}
  {\bibinfo  {journal} {Optics Communications}\ }\textbf {\bibinfo {volume}
  {62}},\ \bibinfo {pages} {374} (\bibinfo {year} {1987})}\BibitemShut
  {NoStop}%
\bibitem [{\citenamefont {Kr{\"o}kel}\ \emph {et~al.}(1988)\citenamefont
  {Kr{\"o}kel}, \citenamefont {Halas}, \citenamefont {Giuliani},\ and\
  \citenamefont {Grischkowsky}}]{krokel_dark-pulse_1988}%
  \BibitemOpen
  \bibfield  {author} {\bibinfo {author} {\bibfnamefont {D.}~\bibnamefont
  {Kr{\"o}kel}}, \bibinfo {author} {\bibfnamefont {N.~J.}\ \bibnamefont
  {Halas}}, \bibinfo {author} {\bibfnamefont {G.}~\bibnamefont {Giuliani}},\
  and\ \bibinfo {author} {\bibfnamefont {D.}~\bibnamefont {Grischkowsky}},\
  }\bibfield  {title} {{\selectlanguage {english}\bibinfo {title} {Dark-{Pulse}
  {Propagation} in {Optical} {Fibers}}},\ }\href
  {https://doi.org/10.1103/PhysRevLett.60.29} {\bibfield  {journal} {\bibinfo
  {journal} {Phys. Rev. Lett.}\ }\textbf {\bibinfo {volume} {60}},\ \bibinfo
  {pages} {29} (\bibinfo {year} {1988})}\BibitemShut {NoStop}%
\bibitem [{\citenamefont {Weiner}\ \emph {et~al.}(1988)\citenamefont {Weiner},
  \citenamefont {Heritage}, \citenamefont {Hawkins}, \citenamefont {Thurston},
  \citenamefont {Kirschner}, \citenamefont {Leaird},\ and\ \citenamefont
  {Tomlinson}}]{weiner_experimental_1988}%
  \BibitemOpen
  \bibfield  {author} {\bibinfo {author} {\bibfnamefont {A.~M.}\ \bibnamefont
  {Weiner}}, \bibinfo {author} {\bibfnamefont {J.~P.}\ \bibnamefont
  {Heritage}}, \bibinfo {author} {\bibfnamefont {R.~J.}\ \bibnamefont
  {Hawkins}}, \bibinfo {author} {\bibfnamefont {R.~N.}\ \bibnamefont
  {Thurston}}, \bibinfo {author} {\bibfnamefont {E.~M.}\ \bibnamefont
  {Kirschner}}, \bibinfo {author} {\bibfnamefont {D.~E.}\ \bibnamefont
  {Leaird}},\ and\ \bibinfo {author} {\bibfnamefont {W.~J.}\ \bibnamefont
  {Tomlinson}},\ }\bibfield  {title} {{\selectlanguage {english}\bibinfo
  {title} {Experimental {Observation} of the {Fundamental} {Dark} {Soliton} in
  {Optical} {Fibers}}},\ }\href {https://doi.org/10.1103/PhysRevLett.61.2445}
  {\bibfield  {journal} {\bibinfo  {journal} {Phys. Rev. Lett.}\ }\textbf
  {\bibinfo {volume} {61}},\ \bibinfo {pages} {2445} (\bibinfo {year}
  {1988})}\BibitemShut {NoStop}%
\bibitem [{\citenamefont {Chong}\ \emph {et~al.}(2013)\citenamefont {Chong},
  \citenamefont {Kevrekidis}, \citenamefont {Theocharis},\ and\ \citenamefont
  {Daraio}}]{chong_dark_2013}%
  \BibitemOpen
  \bibfield  {author} {\bibinfo {author} {\bibfnamefont {C.}~\bibnamefont
  {Chong}}, \bibinfo {author} {\bibfnamefont {P.~G.}\ \bibnamefont
  {Kevrekidis}}, \bibinfo {author} {\bibfnamefont {G.}~\bibnamefont
  {Theocharis}},\ and\ \bibinfo {author} {\bibfnamefont {C.}~\bibnamefont
  {Daraio}},\ }\bibfield  {title} {{\selectlanguage {english}\bibinfo {title}
  {Dark breathers in granular crystals}},\ }\href
  {https://doi.org/10.1103/PhysRevE.87.042202} {\bibfield  {journal} {\bibinfo
  {journal} {Phys. Rev. E}\ }\textbf {\bibinfo {volume} {87}},\ \bibinfo
  {pages} {042202} (\bibinfo {year} {2013})}\BibitemShut {NoStop}%
\bibitem [{\citenamefont {Chong}\ \emph {et~al.}(2014)\citenamefont {Chong},
  \citenamefont {Li}, \citenamefont {Yang}, \citenamefont {Williams},
  \citenamefont {Kevrekidis}, \citenamefont {Kevrekidis},\ and\ \citenamefont
  {Daraio}}]{chong_damped-driven_2014}%
  \BibitemOpen
  \bibfield  {author} {\bibinfo {author} {\bibfnamefont {C.}~\bibnamefont
  {Chong}}, \bibinfo {author} {\bibfnamefont {F.}~\bibnamefont {Li}}, \bibinfo
  {author} {\bibfnamefont {J.}~\bibnamefont {Yang}}, \bibinfo {author}
  {\bibfnamefont {M.~O.}\ \bibnamefont {Williams}}, \bibinfo {author}
  {\bibfnamefont {I.~G.}\ \bibnamefont {Kevrekidis}}, \bibinfo {author}
  {\bibfnamefont {P.~G.}\ \bibnamefont {Kevrekidis}},\ and\ \bibinfo {author}
  {\bibfnamefont {C.}~\bibnamefont {Daraio}},\ }\bibfield  {title}
  {{\selectlanguage {english}\bibinfo {title} {Damped-driven granular chains:
  {An} ideal playground for dark breathers and multibreathers}},\ }\href
  {https://doi.org/10.1103/PhysRevE.89.032924} {\bibfield  {journal} {\bibinfo
  {journal} {Phys. Rev. E}\ }\textbf {\bibinfo {volume} {89}},\ \bibinfo
  {pages} {032924} (\bibinfo {year} {2014})}\BibitemShut {NoStop}%
\bibitem [{\citenamefont {Nesterenko}(2018)}]{nesterenko_waves_2018}%
  \BibitemOpen
  \bibfield  {author} {\bibinfo {author} {\bibfnamefont {V.~F.}\ \bibnamefont
  {Nesterenko}},\ }\bibfield  {title} {{\selectlanguage {english}\bibinfo
  {title} {Waves in strongly nonlinear discrete systems}},\ }\href
  {https://doi.org/10.1098/rsta.2017.0130} {\bibfield  {journal} {\bibinfo
  {journal} {Phil. Trans. R. Soc. A.}\ }\textbf {\bibinfo {volume} {376}},\
  \bibinfo {pages} {20170130} (\bibinfo {year} {2018})}\BibitemShut {NoStop}%
\bibitem [{\citenamefont {Zhang}\ \emph {et~al.}(2018)\citenamefont {Zhang},
  \citenamefont {Romero-Garc{\'i}a}, \citenamefont {Theocharis}, \citenamefont
  {Richoux}, \citenamefont {Achilleos},\ and\ \citenamefont
  {Frantzeskakis}}]{zhang_dark_2018}%
  \BibitemOpen
  \bibfield  {author} {\bibinfo {author} {\bibfnamefont {J.}~\bibnamefont
  {Zhang}}, \bibinfo {author} {\bibfnamefont {V.}~\bibnamefont
  {Romero-Garc{\'i}a}}, \bibinfo {author} {\bibfnamefont {G.}~\bibnamefont
  {Theocharis}}, \bibinfo {author} {\bibfnamefont {O.}~\bibnamefont {Richoux}},
  \bibinfo {author} {\bibfnamefont {V.}~\bibnamefont {Achilleos}},\ and\
  \bibinfo {author} {\bibfnamefont {D.}~\bibnamefont {Frantzeskakis}},\
  }\bibfield  {title} {{\selectlanguage {english}\bibinfo {title} {Dark
  {Solitons} in {Acoustic} {Transmission} {Line} {Metamaterials}}},\ }\href
  {https://doi.org/10.3390/app8071186} {\bibfield  {journal} {\bibinfo
  {journal} {Applied Sciences}\ }\textbf {\bibinfo {volume} {8}},\ \bibinfo
  {pages} {1186} (\bibinfo {year} {2018})}\BibitemShut {NoStop}%
\bibitem [{\citenamefont {Theocharis}\ \emph {et~al.}(2013)\citenamefont
  {Theocharis}, \citenamefont {Boechler},\ and\ \citenamefont
  {Daraio}}]{deymier_nonlinear_2013}%
  \BibitemOpen
  \bibfield  {author} {\bibinfo {author} {\bibfnamefont {G.}~\bibnamefont
  {Theocharis}}, \bibinfo {author} {\bibfnamefont {N.}~\bibnamefont
  {Boechler}},\ and\ \bibinfo {author} {\bibfnamefont {C.}~\bibnamefont
  {Daraio}},\ }\bibfield  {title} {{\selectlanguage {english}\bibinfo {title}
  {Nonlinear {Periodic} {Phononic} {Structures} and {Granular} {Crystals}}},\
  }in\ \href {https://doi.org/10.1007/978-3-642-31232-8_7} {{\selectlanguage
  {english}\emph {\bibinfo {booktitle} {Acoustic {Metamaterials} and {Phononic}
  {Crystals}}}}},\ Vol.\ \bibinfo {volume} {173},\ \bibinfo {editor} {edited
  by\ \bibinfo {editor} {\bibfnamefont {P.~A.}\ \bibnamefont {Deymier}}}\
  (\bibinfo  {publisher} {Springer Berlin Heidelberg},\ \bibinfo {address}
  {Berlin, Heidelberg},\ \bibinfo {year} {2013})\ pp.\ \bibinfo {pages}
  {217--251},\ \bibinfo {note} {series Title: Springer Series in Solid-State
  Sciences}\BibitemShut {NoStop}%
\bibitem [{\citenamefont {Guo}(2018)}]{guo_nonlinear_2018}%
  \BibitemOpen
  \bibfield  {author} {\bibinfo {author} {\bibfnamefont {X.}~\bibnamefont
  {Guo}},\ }{\selectlanguage {english}\emph {\bibinfo {title} {Nonlinear
  architected metasurfaces for acoustic wave scattering manipulation}}},\
  \href@noop {} {\bibinfo {type} {Acoustics [physics.class-ph]}},\ \bibinfo
  {school} {Universit{\'e} du Maine} (\bibinfo {year} {2018}),\ \bibinfo {note}
  {nNT : 2018LEMA1030}\BibitemShut {NoStop}%
\bibitem [{Note1()}]{Note1}%
  \BibitemOpen
  \bibinfo {note} {Note that this model for the elastic potential energy
  assumes that the elastic bonds between vertices behave physically in the
  following way: the bending/rotational restoring moment just depends on the
  relative angles between the neighboring units, the shear restoring force is
  proportional to the elongation of the connector projected on the axis
  orthogonal to the connector axis at rest (e.g. a vertical displacement
  difference of the vertices for a horizontal connector), and the longitudinal
  restoring force is proportional to the elongation of the connector projected
  on the axis of the connector axis at rest. A more general model could be
  implemented, accounting for global rotation effects and geometrical
  nonlinearity associated to large rotations, but would not necessarily lead to
  tractable motion equations. These assumptions have been previously
  experimentally validated for soliton propagation in similar metamaterial
  chains \cite
  {deng_metamaterials_2018,deng_anomalous_2019,deng_nonlinear_2021}.}\BibitemShut
  {Stop}%
\bibitem [{\citenamefont {Kivshar}\ and\ \citenamefont
  {Peyrard}(1992)}]{kivshar_modulational_1992}%
  \BibitemOpen
  \bibfield  {author} {\bibinfo {author} {\bibfnamefont {Y.~S.}\ \bibnamefont
  {Kivshar}}\ and\ \bibinfo {author} {\bibfnamefont {M.}~\bibnamefont
  {Peyrard}},\ }\bibfield  {title} {{\selectlanguage {english}\bibinfo {title}
  {Modulational instabilities in discrete lattices}},\ }\href
  {https://doi.org/10.1103/PhysRevA.46.3198} {\bibfield  {journal} {\bibinfo
  {journal} {Phys. Rev. A}\ }\textbf {\bibinfo {volume} {46}},\ \bibinfo
  {pages} {3198} (\bibinfo {year} {1992})}\BibitemShut {NoStop}%
\bibitem [{\citenamefont {Daumont}\ \emph {et~al.}(1997)\citenamefont
  {Daumont}, \citenamefont {Dauxois},\ and\ \citenamefont
  {Peyrard}}]{daumont_modulational_1997}%
  \BibitemOpen
  \bibfield  {author} {\bibinfo {author} {\bibfnamefont {I.}~\bibnamefont
  {Daumont}}, \bibinfo {author} {\bibfnamefont {T.}~\bibnamefont {Dauxois}},\
  and\ \bibinfo {author} {\bibfnamefont {M.}~\bibnamefont {Peyrard}},\
  }\bibfield  {title} {{\selectlanguage {english}\bibinfo {title} {Modulational
  instability: first step towards energy localization in nonlinear lattices}},\
  }\href {https://doi.org/10.1088/0951-7715/10/3/003} {\bibfield  {journal}
  {\bibinfo  {journal} {Nonlinearity}\ }\textbf {\bibinfo {volume} {10}},\
  \bibinfo {pages} {617} (\bibinfo {year} {1997})}\BibitemShut {NoStop}%
\bibitem [{\citenamefont {Remoissenet}(1986)}]{remoissenet_low-amplitude_1986}%
  \BibitemOpen
  \bibfield  {author} {\bibinfo {author} {\bibfnamefont {M.}~\bibnamefont
  {Remoissenet}},\ }\bibfield  {title} {{\selectlanguage {english}\bibinfo
  {title} {Low-amplitude breather and envelope solitons in
  quasi-one-dimensional physical models}},\ }\href
  {https://doi.org/10.1103/PhysRevB.33.2386} {\bibfield  {journal} {\bibinfo
  {journal} {Phys. Rev. B}\ }\textbf {\bibinfo {volume} {33}},\ \bibinfo
  {pages} {2386} (\bibinfo {year} {1986})}\BibitemShut {NoStop}%
\bibitem [{\citenamefont {Kivshar}\ and\ \citenamefont
  {Agrawal}(2003)}]{kivshar_optical_2003}%
  \BibitemOpen
  \bibfield  {author} {\bibinfo {author} {\bibfnamefont {Y.~S.}\ \bibnamefont
  {Kivshar}}\ and\ \bibinfo {author} {\bibfnamefont {G.}~\bibnamefont
  {Agrawal}},\ }\href@noop {} {{\selectlanguage {english}\emph {\bibinfo
  {title} {Optical {Solitons}: {From} {Fibers} to {Photonic} {Crystals}}}}},\
  \bibinfo {edition} {1st}\ ed.\ (\bibinfo  {publisher} {Academic Press},\
  \bibinfo {year} {2003})\BibitemShut {NoStop}%
\bibitem [{\citenamefont {Remoissenet}(1999)}]{remoissenet_waves_1999}%
  \BibitemOpen
  \bibfield  {author} {\bibinfo {author} {\bibfnamefont {M.}~\bibnamefont
  {Remoissenet}},\ }\href {https://doi.org/10.1007/978-3-662-03790-4}
  {{\selectlanguage {english}\emph {\bibinfo {title} {Waves {Called}
  {Solitons}}}}},\ Advanced {Texts} in {Physics}\ (\bibinfo  {publisher}
  {Springer Berlin Heidelberg},\ \bibinfo {address} {Berlin, Heidelberg},\
  \bibinfo {year} {1999})\BibitemShut {NoStop}%
\bibitem [{\citenamefont {Tikan}\ \emph {et~al.}(2022)\citenamefont {Tikan},
  \citenamefont {Bonnefoy}, \citenamefont {Ducrozet}, \citenamefont
  {Prabhudesai}, \citenamefont {Michel}, \citenamefont {Cazaubiel},
  \citenamefont {Falcon}, \citenamefont {Copie}, \citenamefont {Randoux},\ and\
  \citenamefont {Suret}}]{tikan_nonlinear_2022}%
  \BibitemOpen
  \bibfield  {author} {\bibinfo {author} {\bibfnamefont {A.}~\bibnamefont
  {Tikan}}, \bibinfo {author} {\bibfnamefont {F.}~\bibnamefont {Bonnefoy}},
  \bibinfo {author} {\bibfnamefont {G.}~\bibnamefont {Ducrozet}}, \bibinfo
  {author} {\bibfnamefont {G.}~\bibnamefont {Prabhudesai}}, \bibinfo {author}
  {\bibfnamefont {G.}~\bibnamefont {Michel}}, \bibinfo {author} {\bibfnamefont
  {A.}~\bibnamefont {Cazaubiel}}, \bibinfo {author} {\bibfnamefont
  {{\'E}.}~\bibnamefont {Falcon}}, \bibinfo {author} {\bibfnamefont
  {F.}~\bibnamefont {Copie}}, \bibinfo {author} {\bibfnamefont
  {S.}~\bibnamefont {Randoux}},\ and\ \bibinfo {author} {\bibfnamefont
  {P.}~\bibnamefont {Suret}},\ }\bibfield  {title} {{\selectlanguage
  {english}\bibinfo {title} {Nonlinear dispersion relation in integrable
  turbulence}},\ }\href {https://doi.org/10.1038/s41598-022-14209-7} {\bibfield
   {journal} {\bibinfo  {journal} {Sci Rep}\ }\textbf {\bibinfo {volume}
  {12}},\ \bibinfo {pages} {10386} (\bibinfo {year} {2022})}\BibitemShut
  {NoStop}%
\bibitem [{\citenamefont {Leisman}\ \emph {et~al.}(2019)\citenamefont
  {Leisman}, \citenamefont {Zhou}, \citenamefont {Banks}, \citenamefont
  {Kova{\v c}i{\v c}},\ and\ \citenamefont {Cai}}]{leisman_effective_2019}%
  \BibitemOpen
  \bibfield  {author} {\bibinfo {author} {\bibfnamefont {K.~P.}\ \bibnamefont
  {Leisman}}, \bibinfo {author} {\bibfnamefont {D.}~\bibnamefont {Zhou}},
  \bibinfo {author} {\bibfnamefont {J.~W.}\ \bibnamefont {Banks}}, \bibinfo
  {author} {\bibfnamefont {G.}~\bibnamefont {Kova{\v c}i{\v c}}},\ and\
  \bibinfo {author} {\bibfnamefont {D.}~\bibnamefont {Cai}},\ }\bibfield
  {title} {{\selectlanguage {english}\bibinfo {title} {Effective dispersion in
  the focusing nonlinear {Schr{\"o}dinger} equation}},\ }\href
  {https://doi.org/10.1103/PhysRevE.100.022215} {\bibfield  {journal} {\bibinfo
   {journal} {Phys. Rev. E}\ }\textbf {\bibinfo {volume} {100}},\ \bibinfo
  {pages} {022215} (\bibinfo {year} {2019})}\BibitemShut {NoStop}%
\bibitem [{\citenamefont {Frantzeskakis}(2010)}]{frantzeskakis_dark_2010}%
  \BibitemOpen
  \bibfield  {author} {\bibinfo {author} {\bibfnamefont {D.~J.}\ \bibnamefont
  {Frantzeskakis}},\ }\bibfield  {title} {{\selectlanguage {english}\bibinfo
  {title} {Dark solitons in atomic {Bose}{\textendash}{Einstein} condensates:
  from theory to experiments}},\ }\href
  {https://doi.org/10.1088/1751-8113/43/21/213001} {\bibfield  {journal}
  {\bibinfo  {journal} {J. Phys. A: Math. Theor.}\ }\textbf {\bibinfo {volume}
  {43}},\ \bibinfo {pages} {213001} (\bibinfo {year} {2010})}\BibitemShut
  {NoStop}%
\bibitem [{\citenamefont {Zaera}\ \emph {et~al.}(2018)\citenamefont {Zaera},
  \citenamefont {Vila}, \citenamefont {Fernandez-Saez},\ and\ \citenamefont
  {Ruzzene}}]{zaera_propagation_2018}%
  \BibitemOpen
  \bibfield  {author} {\bibinfo {author} {\bibfnamefont {R.}~\bibnamefont
  {Zaera}}, \bibinfo {author} {\bibfnamefont {J.}~\bibnamefont {Vila}},
  \bibinfo {author} {\bibfnamefont {J.}~\bibnamefont {Fernandez-Saez}},\ and\
  \bibinfo {author} {\bibfnamefont {M.}~\bibnamefont {Ruzzene}},\ }\bibfield
  {title} {{\selectlanguage {english}\bibinfo {title} {Propagation of solitons
  in a two-dimensional nonlinear square lattice}},\ }\href
  {https://doi.org/10.1016/j.ijnonlinmec.2018.08.002} {\bibfield  {journal}
  {\bibinfo  {journal} {International Journal of Non-Linear Mechanics}\
  }\textbf {\bibinfo {volume} {106}},\ \bibinfo {pages} {188} (\bibinfo {year}
  {2018})}\BibitemShut {NoStop}%
\bibitem [{\citenamefont {Tikan}\ \emph {et~al.}(2017)\citenamefont {Tikan},
  \citenamefont {Billet}, \citenamefont {El}, \citenamefont {Tovbis},
  \citenamefont {Bertola}, \citenamefont {Sylvestre}, \citenamefont {Gustave},
  \citenamefont {Randoux}, \citenamefont {Genty}, \citenamefont {Suret},\ and\
  \citenamefont {Dudley}}]{tikan_universality_2017}%
  \BibitemOpen
  \bibfield  {author} {\bibinfo {author} {\bibfnamefont {A.}~\bibnamefont
  {Tikan}}, \bibinfo {author} {\bibfnamefont {C.}~\bibnamefont {Billet}},
  \bibinfo {author} {\bibfnamefont {G.}~\bibnamefont {El}}, \bibinfo {author}
  {\bibfnamefont {A.}~\bibnamefont {Tovbis}}, \bibinfo {author} {\bibfnamefont
  {M.}~\bibnamefont {Bertola}}, \bibinfo {author} {\bibfnamefont
  {T.}~\bibnamefont {Sylvestre}}, \bibinfo {author} {\bibfnamefont
  {F.}~\bibnamefont {Gustave}}, \bibinfo {author} {\bibfnamefont
  {S.}~\bibnamefont {Randoux}}, \bibinfo {author} {\bibfnamefont
  {G.}~\bibnamefont {Genty}}, \bibinfo {author} {\bibfnamefont
  {P.}~\bibnamefont {Suret}},\ and\ \bibinfo {author} {\bibfnamefont {J.~M.}\
  \bibnamefont {Dudley}},\ }\bibfield  {title} {{\selectlanguage
  {english}\bibinfo {title} {Universality of the {Peregrine} {Soliton} in the
  {Focusing} {Dynamics} of the {Cubic} {Nonlinear} {Schr{\"o}dinger}
  {Equation}}},\ }\href {https://doi.org/10.1103/PhysRevLett.119.033901}
  {\bibfield  {journal} {\bibinfo  {journal} {Phys. Rev. Lett.}\ }\textbf
  {\bibinfo {volume} {119}},\ \bibinfo {pages} {033901} (\bibinfo {year}
  {2017})}\BibitemShut {NoStop}%
\bibitem [{\citenamefont {Charalampidis}\ \emph {et~al.}(2018)\citenamefont
  {Charalampidis}, \citenamefont {Lee}, \citenamefont {Kevrekidis},\ and\
  \citenamefont {Chong}}]{charalampidis_phononic_2018}%
  \BibitemOpen
  \bibfield  {author} {\bibinfo {author} {\bibfnamefont {E.~G.}\ \bibnamefont
  {Charalampidis}}, \bibinfo {author} {\bibfnamefont {J.}~\bibnamefont {Lee}},
  \bibinfo {author} {\bibfnamefont {P.~G.}\ \bibnamefont {Kevrekidis}},\ and\
  \bibinfo {author} {\bibfnamefont {C.}~\bibnamefont {Chong}},\ }\bibfield
  {title} {{\selectlanguage {english}\bibinfo {title} {Phononic rogue waves}},\
  }\href {https://doi.org/10.1103/PhysRevE.98.032903} {\bibfield  {journal}
  {\bibinfo  {journal} {Phys. Rev. E}\ }\textbf {\bibinfo {volume} {98}},\
  \bibinfo {pages} {032903} (\bibinfo {year} {2018})}\BibitemShut {NoStop}%
\bibitem [{\citenamefont {Tikan}\ \emph {et~al.}(2021)\citenamefont {Tikan},
  \citenamefont {Randoux}, \citenamefont {El}, \citenamefont {Tovbis},
  \citenamefont {Copie},\ and\ \citenamefont {Suret}}]{tikan_local_2021}%
  \BibitemOpen
  \bibfield  {author} {\bibinfo {author} {\bibfnamefont {A.}~\bibnamefont
  {Tikan}}, \bibinfo {author} {\bibfnamefont {S.}~\bibnamefont {Randoux}},
  \bibinfo {author} {\bibfnamefont {G.}~\bibnamefont {El}}, \bibinfo {author}
  {\bibfnamefont {A.}~\bibnamefont {Tovbis}}, \bibinfo {author} {\bibfnamefont
  {F.}~\bibnamefont {Copie}},\ and\ \bibinfo {author} {\bibfnamefont
  {P.}~\bibnamefont {Suret}},\ }\bibfield  {title} {{\selectlanguage
  {english}\bibinfo {title} {Local {Emergence} of {Peregrine} {Solitons}:
  {Experiments} and {Theory}}},\ }\href
  {https://doi.org/10.3389/fphy.2020.599435} {\bibfield  {journal} {\bibinfo
  {journal} {Front. Phys.}\ }\textbf {\bibinfo {volume} {8}},\ \bibinfo {pages}
  {599435} (\bibinfo {year} {2021})}\BibitemShut {NoStop}%
\end{thebibliography}%

%

\end{document}